\documentclass[12pt,a4paper]{iopart}

\PassOptionsToPackage{numbers, sort&compress}{natbib}

\usepackage[utf8]{inputenc} 
\usepackage[T1]{fontenc}    
\usepackage{hyperref}       
\usepackage{url}            
\usepackage{booktabs}       
\usepackage{amsfonts}       
\usepackage{nicefrac}       
\usepackage{microtype}      
\usepackage{pdflscape}
\usepackage{subfigure}
\usepackage{cite}
\usepackage{IEEEtrantools}
\usepackage{graphicx}
\usepackage{placeins}
\usepackage{lscape}

\newcommand{\orcid}[1]{\href{https://orcid.org/#1}{\,\includegraphics[width=8px]{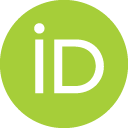}}}

\begin{document}

\title[Expressive Gaussian mixture models for high-dimensional statistical modelling]{Learning to discover: expressive Gaussian mixture models for multi-dimensional simulation and parameter inference in the physical sciences}

\author{Stephen B.\ Menary\orcid{0000-0003-1244-2802} \,\& Darren D.\ Price\orcid{0000-0003-2750-9977}}
\address{Department of Physics \& Astronomy,\\ University of Manchester, UK}
\ead{stephen.menary@manchester.ac.uk, darren.price@manchester.ac.uk}

\begin{abstract}
  We show that density models describing multiple observables with (i) hard boundaries and (ii) dependence on external parameters may be created using an auto-regressive Gaussian mixture model. The model is designed to capture how observable spectra are deformed by hypothesis variations, and is made more expressive by projecting data onto a configurable latent space. It may be used as a statistical model for scientific discovery in interpreting experimental observations, for example when constraining the parameters of a physical model or tuning simulation parameters according to calibration data. The model may also be sampled for use within a Monte Carlo simulation chain, or used to estimate likelihood ratios for event classification. The method is demonstrated on simulated high-energy particle physics data considering the anomalous electroweak production of a $Z$ boson in association with a dijet system at the Large Hadron Collider, and the accuracy of inference is tested using a realistic toy example. The developed methods are domain agnostic; they may be used within any field to perform simulation or inference where a dataset consisting of many real-valued observables has conditional dependence on external parameters.
\end{abstract}

\noindent{\it Keywords\/}: statistical inference, scientific discovery, machine learning, simulation

\section{Introduction}
\label{sec:intro}

In the physical sciences, we often use statistical methods to make quantifiable statements about how compatible experimental observations are with different hypotheses about nature. These frameworks, typically frequentist or Bayesian, usually require us to model the expected probability density function (PDF) for all possible observations, conditioned on the hypotheses of interest. Finding such a parameterization for the PDF can be very challenging when data are multi-dimensional.

Within experimental particle physics, often the problem is simplified by observing only one or two dimensions of the data at a time following some initial data selections. For these low-dimensional measurements, we are able to approximate the PDF either parametrically or using histograms, allowing for statistical interpretation of the data. To ensure these simplified measurements contain maximum sensitivity to the processes of interest, hereafter referred to as the ``signal'' in contrast with the ``background'' of all other processes contained in the dataset, we only select data in regions of phase space for which the frequency of signal is high relative to the background. We note several disadvantages of this approach:
\begin{enumerate}
    \item By analyzing data only in select regions of phase space, we lose any potentially useful information contained within other regions.
    \item Different hypotheses may predict different distributions of the data in the high-dimensional space. However, we lose this information when collapsing data into one or two dimensions.
    \item When analyzing histograms, the binning of data discards finely-grained information about the shape of the distribution. 
    \item The experimentalist must manually design the selection criteria, observables and binning, making it difficult to ensure that an analysis provides fully optimized sensitivity to all accessible regions of the theory parameter space.
\end{enumerate}
If the expected signal and background PDFs can be modeled parametrically in a space spanning all data dimensions, the PDF ratio contains the expected signal-to-background-ratio at every point in phase space. This means that we do not require restrictive data selections to optimize statistical sensitivity to the signal component. We also do not require binning. The information described above is therefore retained and may be used to provide greater exclusion and discovery potential for all possible new physics models.\footnote{Here we consider only the optimization of statistical sensitivity and assume that the PDFs can be modeled with sufficient accuracy and well-described systematic uncertainties. This may be challenging in a real-world analysis which includes data-driven constraints and regions with large systematic effects.}

It has recently been demonstrated that machine-learned density models may be constructed which describe PDFs (or PDF ratios) in a high-dimensional observable space \cite{Brehmer:2018eca,Brehmer:2019xox,Brehmer5242,cranmer2016approximating,papamakarios2018masked,DBLP:journals/corr/UriaCGML16,alsing-lfi-cosmo,dinh2017density,_t_p_nek_2015}. Provided that model bias can be mitigated and systematic uncertainties properly described, these can be used to perform parameter inference or construct likelihood ratios for event classification.\footnote{See e.g. Refs~\cite{barron2021unsupervised,PhysRevD.100.035040,kasieczka2021lhc} for alternative approaches for enhancing sensitivity to new physics models using machine-learned classifiers and anomaly detection.}

Many PDF models may also be sampled from, which is not the case when exclusively modeling the PDF ratio. This has several benefits:
\begin{enumerate}
    \item We can verify that the distribution obtained by sampling the model is well-behaved when compared with the training data. Such cross checks are desirable in the physical sciences, where rigorous data interpretation is emphasized.
    \item It may be used to generate new datasets at arbitrary points in parameter space, which the model accomplishes by interpolating between the external parameter values at which training data were provided.
    \item We can numerically estimate the expected distribution of a test-statistic under different parameter hypotheses, instead of assuming an asymptotic form. This aids in the estimation of rigorous frequentist confidence limits.
    \item Once trained, sampling from the density model may be more computationally efficient than running the full simulation package used to generate training data. In this context, density models provide a compelling alternative to other stochastic generative models such as generative adversarial networks (GANs)\,\cite{goodfellow2014generative} and variational auto-encoders (VAEs)\,\cite{kingma2014autoencoding,Kingma_2019} for performing steps in a simulation chain\,\cite{dijetgan_2019,butter_2019,butter2020generative,ATL-SOFT-PUB-2018-001}.
\end{enumerate} 

In this work, we will show that density models describing multiple observables with (i) a complex multi-dimensional distribution, (ii) hard boundaries and (iii) dependence on external parameters may be created using an \textit{auto-regressive Gaussian mixture model} \cite{Bishop_1994,Variani_2015,DBLP:journals/corr/UriaCGML16,papamakarios2018masked}\footnote{Whilst we were unable to find examples which combine all these properties, Refs~\cite{Bishop_1994,Variani_2015} provide examples of Gaussian mixture models for density estimation parameterized using neural networks and Refs~\cite{DBLP:journals/corr/UriaCGML16,papamakarios2018masked} of auto-regressive density estimation used to model multi-dimensional data.}. The model is made more expressive by projecting data onto a configurable latent space. The method is designed to capture how observable spectra are continuously deformed as the external parameters are varied, behavior which is common in the physical sciences. We hope that this work will provide users with a simple but expressive way to model such datasets in their own domains.

To study the performance of our method on a high-dimensional dataset of physically realistic observables, we use simulations of particle physics data sensitive to anomalies in the electroweak production of a $Z$ boson in association with a dijet system. We demonstrate the degree to which our trained density models can describe this data, capturing how it is deformed as two physical parameters are varied. We then use a toy example, in which we can access the ground-truth PDF, to demonstrate that accurate parameter estimates and exclusion limits may be obtained using our method. This is not possible using the physical example because we do not have access to the ground-truth PDF with which to compare. 

This paper is structured as follows. In Section~\ref{sec:exp} we describe the generation of training data used throughout the paper, and explain the physical basis behind it. In Section~\ref{sec:method} we describe how data are transformed onto the latent space and how the density model is built. We then discuss several features of the model. In Section~\ref{sec:VBFZ-11obs} we construct a $12$-dimensional model to study the ability to describe a highly multi-dimensional dataset. In Section~\ref{sec:VBFZ-4obs-2param} we construct a $4$-dimensional model with dependence on two external parameters to study the ability to learn the parameter dependence. In Section~\ref{sec:toy-model-inference} we study the accuracy of inference using our toy example. In Section~\ref{sec:conclusion} we conclude.

Whilst these experiments demonstrate that the method is performant on datasets of realistic observables within the domain of high-energy physics, we emphasize that it may be used to model any dataset of continuous observables for which a high-dimensional PDF is deformed by parameter variations, regardless of scientific domain, provided that appropriate training data may be provided.

\section{Experimental setup}
\label{sec:exp}

To test our method in a real-world environment, we consider the electroweak production of a $Z$ boson in association with a dijet system occurring in high-energy proton--proton collisions at the Large Hadron Collider. This process is labeled EW $Zjj$ in the remainder of this text. It is often referred to as the Vector Boson Fusion production of a $Z$ boson.

We choose to model the EW $Zjj$ process for several reasons. Firstly, it provides a number of physically interesting observables which are correlated, challenging our method to capture a feature-rich high-dimensional distribution. Secondly, there exist new physics models which are expected to continuously deform this distribution in distinct ways as different parameters-of-interest are varied. Finally, it is a process of interest for current and future LHC experiments. Nonetheless, we emphasize that the EW $Zjj$ process is intended to be a representative example using which we test the ability of our method to overcome general modeling challenges, and we hope that the method may be used to model smoothly-varying parameter-dependent high-dimensional datasets in any domain.

Each `event' is the observation of many particles created by a single proton--proton collision. High-energy physics datasets typically consists of $\mathcal{O}\left(100-100\mathrm{M}\right)$ events, depending on the pre-selection criteria applied. By identifying the particles produced, and measuring their kinematic properties and other high-level `observables', we study the processes which contributed to their production.

The EW $Zjj$ process is characterized by a final state of two jets of hadrons along with two oppositely charged electrons or muons which are produced by a $Z$-boson decay. Since the EW $Zjj$ process is defined by a $t$-channel exchange of a colour-neutral weak boson between the two incoming partons, these jets are typically separated by a wider rapidity than in the dominant background process which contains a $t$-channel exchange of a gluon. As a result, experimental analyses often select events with a large dijet rapidity separation (or large invariant mass) to enhance the proportion of signal within their sample. We may measure the event rate as a function of many observables. We expect that the presence of certain new particles/forces will induce distortions in the shape or magnitude of these spectra relative to the precise predictions of the Standard Model of Particle Physics (SM).  These measurements enable a rich discovery potential for new natural phenomena and the derivation of constraints on the theoretical models describing them.

The binned one-dimensional kinematic spectra of particles produced via EW $Zjj$ in high-energy proton--proton collisions were recently measured\,\cite{atlascollaboration2020differential,ATLAS:2017nei} by the ATLAS experiment\,\cite{ATLAS:2008xda}. Exclusion limits were derived for several parameters of the Standard Model (SM) effective field theory (SMEFT) in the Warsaw basis\,\cite{Grzadkowski_2010}, which characterize the presence of any novel physics phenomena in such interactions. In this work, we consider how EW $Zjj$ events are affected by variations of the SMEFT parameters $c_{\rm HWB}$ and ${\tilde c}_W$. These parameters extend the SM Lagrangian $\mathcal{L}_{\rm SM}$ by the addition of two non-renormalizable terms with mass dimension six. These additional terms modify how electroweak bosons interact with one another, impacting the rate and expected kinematic distribution of EW $Zjj$ events. These modifications reflect the indirect effects of new physics interactions above some energy scale $\Lambda$ which is not directly probed by the experiment. We will assume $\Lambda = 1~{\rm TeV}$ throughout, noting that other choices simply correspond to a re-scaling of $c_{\rm HWB}$ and ${\tilde c}_W$ within this parameterization. The effective Lagrangian is \cite{Grzadkowski_2010,Brivio_2019,Ellis_2018}
\begin{equation}
    \mathcal{L} ~=~ \mathcal{L}_{\rm SM} ~+~ \frac{c_{\rm HWB}}{\Lambda^2} H^\dagger \tau^I H W_{\mu\nu}^I B^{\mu\nu} ~+~ \frac{{\tilde c}_{W}}{\Lambda^2} \epsilon^{IJK} {\tilde W}^{I\nu}_\mu W^{J\rho}_\nu W^{K\mu}_\rho
\end{equation}
where $H$ is the Higgs doublet, $\tau$ are the Pauli matrices, $W^{\mu\nu}$ and $B^{\mu\nu}$ are the electroweak field strength tensors, $\epsilon$ are anti-symmetric tensors with $\epsilon_{012} = \epsilon_{0123}=1$, ${\tilde W}^{\mu\nu} = \frac{1}{2}\epsilon^{\mu\nu}_{\rho\sigma}W^{\rho\sigma}$ and we neglect Hermitian conjugates. In this work, we use simulated events to construct high-dimensional statistical models which describe many of the kinematic observables considered in the ATLAS analysis. Of the six parameters constrained within the ATLAS analysis, we choose to study $c_{\rm HWB}$ and ${\tilde c}_W$ because they are shown to vary the expected PDF in distinctly different ways. Simultaneously modeling both parameters therefore provides a more ambitious test for the efficacy of our methods.

Ground truth events are generated using the \texttt{Madgraph5} (\texttt{MG5})\,\cite{Alwall:2014hca} program with perturbative calculations at leading order in the strong coupling constant. This models the primary high-energy interaction of interest, simulating the resultant array of particles and their properties. Subsequent hadronization of these particles and modeling of the underlying event\,\cite{Buckley:2011ms,Zyla:2020zbs} are simulated using \texttt{Pythia8}\,\cite{Sjostrand:2014zea,Sjostrand:2007gs}. Definition and selection of stable and detectable particles produced in the collision is performed using \texttt{Rivet}\,\cite{Bierlich:2019rhm}. Neural networks are implemented using \texttt{TensorFlow} v2.4.3 interfaced with \texttt{Keras} v2.4.0\,\cite{tensorflow2015-whitepaper,chollet2015keras}. SMEFT interactions are implemented in \texttt{MG5} using the \texttt{SMEFTSim} \cite{Brivio_2017} package. $1$M datapoints are generated at the Standard Model value of $\left(c_{\rm HWB},~{\tilde c}_W\right) = \left(0,0\right)$. $400$k datapoints are generated in increments of $0.1$ on the interval ${\tilde c}_W \in \left[-0.4,~0.4\right]$ with $c_{\rm HWB}=0$, excluding the SM configuration. $200$k datapoints are generated in a 2D grid with increments of $0.2$ on the interval ${\tilde c}_W \in \left[-0.4,~0.4\right]$ and increments of $2$ on the interval $c_{\rm HWB} \in \left[-4,~4\right]$, excluding pairs with $c_{\rm HWB}=0$.

All objects are defined at particle level, i.e.\ after parton showering and hadronization (as they would appear in a particle detector). Testing our method on such a dataset demonstrates that it fulfills the key objective of this work: to effectively model a high-dimensional PDF of physically realistic observables with external parameter dependence. Since the method is not restricted to any particular experiment or domain, we do not simulate the effects of detector efficiency and resolution when generating our training data. However, we note that end-users who wish to perform (for example) parameter estimation using detector-level experimental data can accomplish this by simulating the impact of their detector when generating their own training data. We expect this to smear the PDF, but not impact the key modeling challenges identified above. We emphasize that there are no practical barriers preventing the modeling of detector-level datasets for use within a given experimental context.

\section*{EW $Zjj$ event selection and observable definitions}

Selection requirements and observables of interest are chosen based on the recent ATLAS measurement\,\cite{atlascollaboration2020differential}, and the ATLAS co-ordinate system\,\cite{ATLAS:2008xda} is used throughout with all observables defined in the laboratory reference frame.

All final state objects are required to satisfy a pseudorapidity of $|\eta|\leq5$. Electrons and muons are `dressed'\,\cite{ATL-PHYS-PUB-2015-013} with photons within a cone of $\Delta R \leq 0.1$. Electrons are required to satisfy $p_{\rm T} \geq 25~{\rm GeV}$ and have $|\eta|<2.47$ excluding $1.37 < |\eta| < 1.52$ where $p_{\rm T}$ is the momentum component transverse to the beamline. Muons are required to satisfy $p_{\rm T} \geq 25~{\rm GeV}$ and $|\eta|<2.4$. Jets arise from collimated streams of stable particles and are clustered\,\cite{Cacciari:2011ma} from all final state particles excluding muons and neutrinos using the anti-$k_{\rm T}$ algorithm\,\cite{Cacciari:2008gp} within a cone of $\Delta R \leq 0.4$. Reconstructed jets are required to satisfy $p_{\rm T} \geq 30~{\rm GeV}$ and have a rapidity of $|y|<4.4$. Jets are rejected if they fall within $\Delta R \leq 0.2$ of a selected electron, to reflect the limitations of a real detector in accurately distinguishing jets and electrons produced at small angular separations.

Events are required to have at least two selected electrons or muons, where the two leptons with the highest $p_{\rm T}$ are used to define the dilepton system and are required to have opposite charge. Events are also required to contain two selected jets, and the two jets with the highest $p_{\rm T}$ are used to define the dijet system. The following observables are calculated from the selected objects:
\begin{itemize}
    \item $m_{\rm ll}$, $p_{\rm T}^{\rm ll}$ and $|y^{\rm ll}|$ are respectively the mass, transverse momentum and absolute rapidity of the dilepton system.
    \item $m_{\rm jj}$, $p_{\rm T}^{\rm jj}$ and $|y^{\rm jj}|$ are respectively the mass, transverse momentum and absolute rapidity of the dijet system.
    \item $p_{\rm T}^{\rm j1}$ and $p_{\rm T}^{\rm j2}$ are the transverse momenta of the highest and second-highest $p_{\rm T}$ jets.
    \item $\Delta \phi \left(j,j\right)$ is the angular spread of the dijet system in a plane transverse to the beamline, measured clockwise with respect to the highest rapidity jet and defined on a domain of $[-\pi,~\pi]$.
    \item $|\Delta y \left(j,j\right)|$ is the absolute rapidity spread of the dijet system.
    \item $N_{\rm jet}$ is the number of selected jets, and $N_{\rm gapjet}$ is the number of selected jets which have a rapidity in the interval bounded by the rapidities of the two highest $p_{\rm T}$ jets.
\end{itemize}
Table~\ref{tab:obs-lims} shows the intervals over which these observables are defined. Events are rejected if any observable falls outside of its interval. The total selection efficiency is estimated to be 64~\% using the events simulated under the SM hypothesis.

\begin{table}[htbp]
\caption{Closed intervals over which observables are selected for experiments performed on simulated EW $Zjj$ data. Events are rejected if they fail any selection requirement.}
\label{tab:obs-lims}
\begin{indented}
\item[]\begin{tabular}{ll}
\br
        Observable   &   Closed interval  \\
\mr
         $m_{\rm ll}$          &  $[75,~105]$ GeV  \\
         $p_{\rm T}^{\rm ll}$  &  $[0,~900]$ GeV   \\
         $y^{\rm ll}$          &  $[0,~2.2]$   \\
         $m_{\rm jj}$          &  $[150,~5000]$ GeV  \\
         $p_{\rm T}^{\rm jj}$  &  $[0,~900]$ GeV   \\
         $y^{\rm jj}$          &  $[0,~4.4]$   \\
         $p_{\rm T}^{\rm j1}$  &  $[60,~1200]$ GeV   \\
         $p_{\rm T}^{\rm j2}$  &  $[40,~1200]$ GeV   \\
         $\Delta \phi \left(j,j\right)$  &  $[-\pi, \pi]$   \\
         $|\Delta y \left(j,j\right)|$  &  $[0,~8.8]$   \\
         $N_{\rm jet}$  &  $[0,~5]$   \\
         $N_{\rm gapjet}$  &  $[0,~2]$   \\
\br
    \end{tabular}
\end{indented}
\end{table}

\section{Method overview}
\label{sec:method}

Consider that we measure datapoints $x \in \mathbb{X}$ on an $n$-dimensional observable space $\mathbb{X} \equiv \mathbb{R}^n$. The PDF is $p(x|\theta)$, where $\theta \in \Theta$ represents the set of parameters of interest and nuisance parameters. This conditional dependence allows us to constrain a set of possible physical models according to their consistency with experimental observations. 

\section*{Gaussian mixture models}

We can model a conditional \textit{one-dimensional} density $p(x|\theta)$ by simulating data for a variety of $\theta$ and fitting this with a conditional Gaussian mixture model (GMM). This parameterizes the density as a linear sum of Gaussian distributions according to
\begin{equation}
    p_{\phi}\left(x|\theta\right) = \sum_{g=1}^{N_G} f_{\phi, g}\left(\theta\right) \cdot \mathcal{N}\left(x;~\mu_{\phi, g}\left(\theta\right);~\sigma_{\phi, g}\left(\theta\right) \right)
\end{equation}
where $N_G$ labels the number of Gaussian modes; $\mathcal{N}$ is a Gaussian probability density function; $f_{\phi, g}$, $\mu_{\phi, g}$ and $\sigma_{\phi, g}$ are respectively the amplitude, mean and width of the $g^{\rm th}$ Gaussian subject to $\sum_{g=1}^{N_G}f_{\phi, g}=1$ and $f_{\phi, g} \geq 0 ~\forall~ g$; $\phi$ label the parameters of a neural network used to capture the functional forms of $f_{\phi, g}$, $\mu_{\phi, g}$ and $\sigma_{\phi, g}$ (see e.g. \cite{Bishop_1994,Variani_2015}).

We use mixture models in this work because they allow us to model arbitrarily complex positive-definite distributions which can be analytically normalized to unity and easily sampled from. This is achieved by writing the density as the linear sum of simple parametric probability distributions. They are often used to model multi-modal data \cite{doi:10.1146/annurev-statistics-031017-100325}, and are well-suited for our probability spectra which we can imagine as being composed from a series of overlapping local probability masses. Each local mass may be modeled as having a different dependence on the external parameters $\theta$, allowing us to express how every region of the spectrum is deformed when $\theta$ is varied. In this work we use Gaussian distributions to model each local mass of density. This is because they are simple distributions (each defined by only two parameters) which are peaked in the center and smoothly vary to $0$ without excessively sharp or sparse tails, ensuring continuity in the model and retaining the local nature of the probability mass. They are also easily normalized and sampled from.

However, there are several ways in which the shape of $p(x|\theta)$ may not be well-suited to a GMM:
\begin{enumerate}
    \item GMMs naturally model a smooth turn-off at the boundaries of a distribution, whereas the data distribution may have hard boundaries due to strict physical constraints or event pre-selection.
    \item The structural features of the PDF, and any deformations induced by variations of $\theta$, must be smooth and wide enough to be modulated by the Gaussian modes.
    \item In order to deform the PDF \textit{downwards}, the model must contain a Gaussian mode with finite amplitude local to the deformation, the amplitude of which can be modulated downwards without impacting the rest of the distribution.\footnote{A density model must be positive definite everywhere. For a GMM, we enforce this by only allowing positive amplitudes for the Gaussian modes. To deform the PDF upwards in some local region, we can add a new Gaussian mode with positive amplitude. However, a downwards deformation cannot be similarly accounted for by adding a new Gaussian mode with negative amplitude, as this is not allowed. This can only be described if the nominal model already contained a narrow Gaussian mode with positive amplitude local to the deformation. In this case we capture the downwards deformation by modulating the amplitude downwards, from positive to less-positive.}
\end{enumerate}

Points (ii) and (iii) mean that a GMM which is dominated by few wide Gaussian modes will have limited ability to describe local deformations of the PDF as $\theta$ is varied. Instead, we wish to have a distribution which is described by \textit{a spectrum of many narrow overlapping Gaussian modes} and which contains \textit{no deformations narrower than the Gaussians themselves}. We will now show that these conditions may be achieved by transforming the data and using a suitable network architecture to model $f_{\phi, g}$, $\mu_{\phi, g}$ and $\sigma_{\phi, g}$. We find that this method resolves the failure conditions listed above in the experiments presented.

\section*{Modeling a single observable}

Datapoints are projected by a function $h : x \mapsto u \in \mathbb{U}$ onto a latent space $\mathbb{U} \equiv \mathbb{R}^n$. The properties of the projection may be tuned to optimize the performance of a GMM describing the density $p_\phi(u|\theta)$. We will now explore this idea using our EW $Zjj$ example.

\begin{figure}[bhtp]
    \centering
    \includegraphics[width=0.95\textwidth]{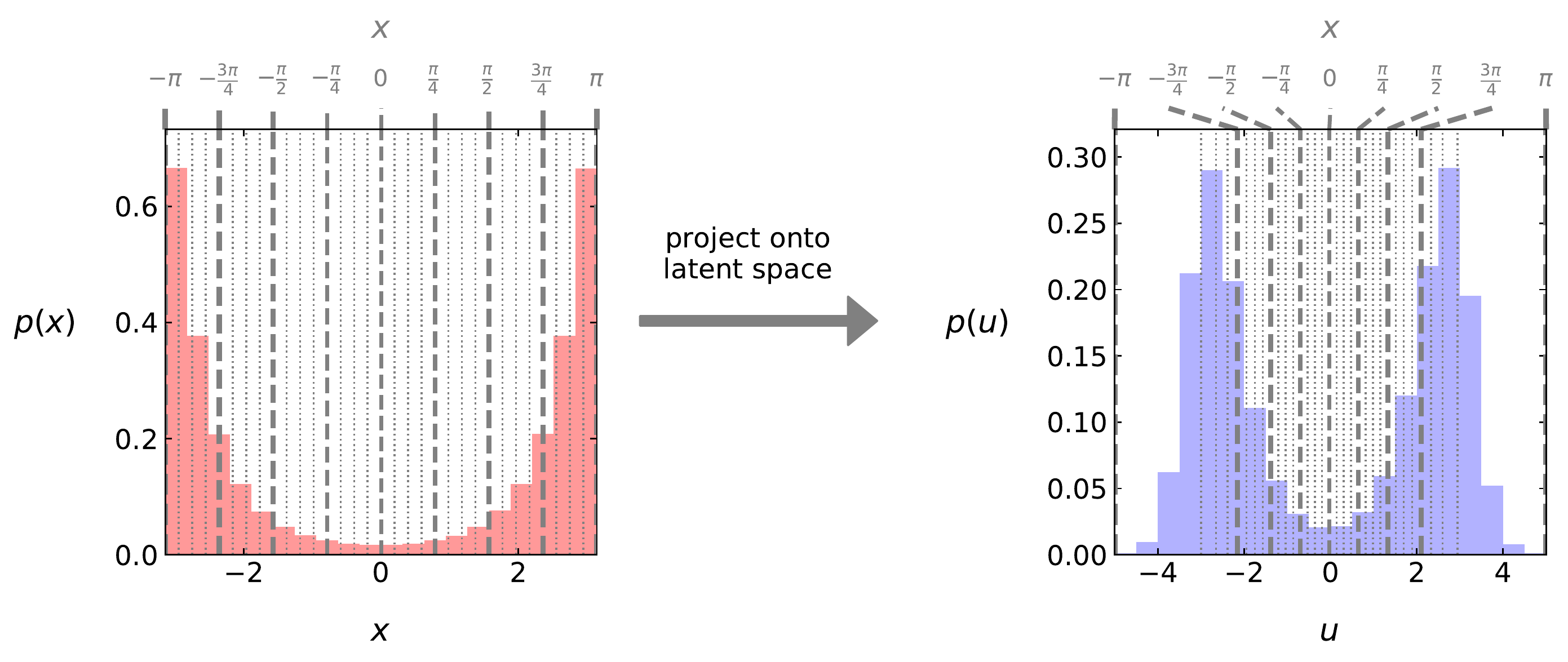}
    \caption{Left: probability density $p\left(x\right)$ with $x=\Delta\phi\left(j,j\right)$, evaluated using \texttt{MG5} events assuming $c_{\rm HWB}={\tilde c}_W=0$. Right: probability density after projecting onto the the latent space using the method described in Figure~\ref{fig:proj-dists-2}.}
    \label{fig:proj-dists-1}
\end{figure}

Consider the case where $x = \Delta\phi\left(j,j\right)$ is the only observable. Figure~\ref{fig:proj-dists-1} (left) shows the probability density $p\left(x\right)$ for the SM case of $c_{\rm HWB}={\tilde c}_W=0$. This plot is obtained by histogramming the datapoints simulated using \texttt{MG5}. We note that this distribution has hard physical boundaries at $[-\pi,~\pi]$ which a GMM would be unable to model. Figure~\ref{fig:proj-dists-1} (right) shows the probability density of the same datapoints after projecting $x$ onto the latent space. This distribution is designed to be well described by a series of overlapping narrow Gaussian modes. We will now describe how this projection function $h\left(x\right)$ was derived, then train a GMM to model this spectrum for a variety of ${\tilde c}_W$.

\begin{figure}[bhtp]
    \centering
    \includegraphics[width=0.99\textwidth]{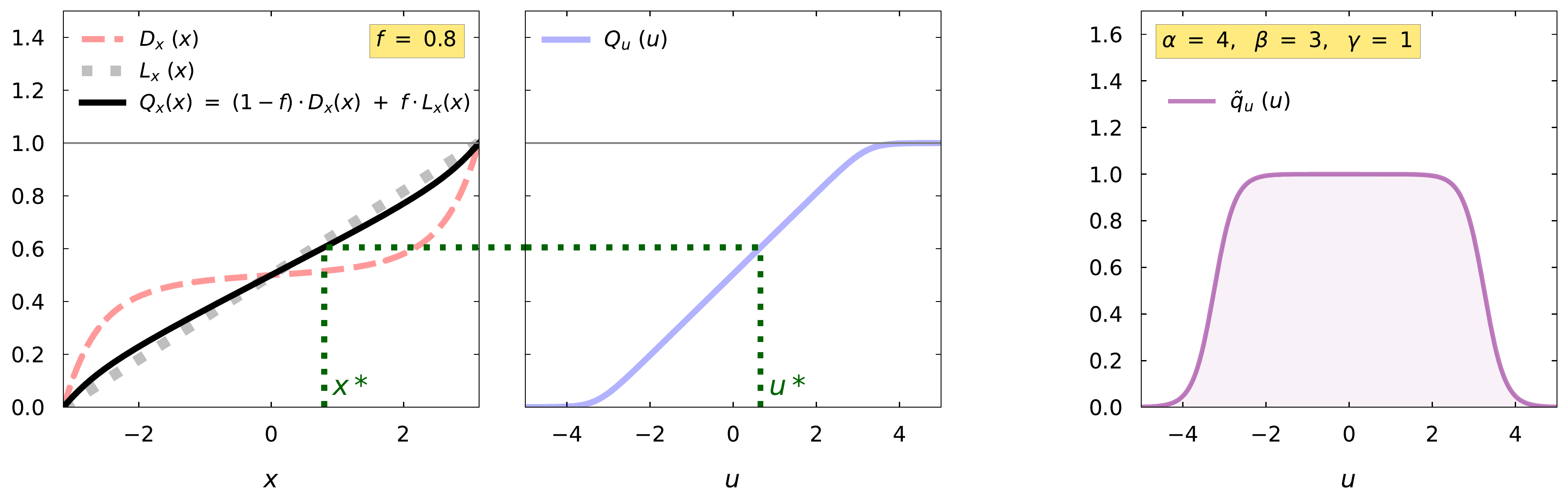}
    \caption{Left: response curve over the data space, $Q_x\left(x\right)$, derived as the linear sum of $D_x\left(x\right)$ and $L_x\left(x\right)$. Middle: response curve over the latent space, $Q_u\left(u\right)$, derived as the cumulative distribution function of ${\tilde q}_u\left(u\right)$. Right: heuristic function ${\tilde q}_u\left(u\right)$. The green dotted line connecting $Q_x\left(x\right)$ with $Q_u\left(u\right)$ visually represents how a datapoint at $x^*$ is transformed onto $u^*$ in the latent space.}
    \label{fig:proj-dists-2}
\end{figure}

To derive $h\left(x\right)$, we first construct a response curve $Q_x\left(x\right)$ between the physical boundaries of $x$. This is written as
\begin{equation}
    Q_x\left(x\right) = \left(1-f\right)\cdot D_x\left(x\right) + f\cdot L_x\left(x\right)
\end{equation}
where $D_x\left(x\right)$ is the cumulative distribution function of the data simulated at the SM and $L_x\left(x\right)$ is a linear function. The hyperparameter $f$ is tuned to ensure that wide regions in $\mathbb{X}$ are not collapsed onto narrow regions in $\mathbb{U}$, whilst also providing a smooth turn-off at the boundaries of the distribution. This function is shown as the solid black line in Figure~\ref{fig:proj-dists-2} (left). We then construct a response curve $Q_u\left(u\right)$ over the latent space, shown as the solid blue line in Figure~\ref{fig:proj-dists-2} (middle), defined as the cumulative distribution function of a target function ${\tilde q}_u\left(u\right)$ given by
\begin{equation}
{\tilde q}_u\left(u\right) = \frac{1}{1+\exp[\alpha(u-\beta)-\gamma]} \cdot \frac{1}{1+\exp[-\alpha(u+\beta)-\gamma]} ~~.
\label{Eq::map-func}
\end{equation}
This function, shown in Figure~\ref{fig:proj-dists-2} (right) using values of $\left(\alpha, \beta,\gamma\right)=\left(4,3,1\right)$, is heuristically designed to be flat in the centre and smooth at the edges. This encourages the optimal GMM description to contain many narrow overlapping Gaussian modes. We note that it may seem natural to choose a Gaussian distribution for ${\tilde q}_u\left(u\right)$ (see e.g.\,\cite{_t_p_nek_2015}), however this will often result in a GMM which is dominated by a single wide Gaussian mode, violating our target behaviour. The mapping function between $\mathbb{X}$ and $\mathbb{U}$ is defined as $h\left(x\right) = Q_u^{-1}\left(Q_x\left(x\right)\right)$, and its derivation is shown visually as the green dotted line connecting the points $x^*$ and $u^*$ in Figure~\ref{fig:proj-dists-2} (left and middle).

We compute $Q_u\left(u\right)$ as a piecewise-linear function over the interval $u\in\left[-5,~5\right]$. Whilst the domain of $u$ could be extended arbitrarily far so that all sampled points $u^*\in\mathbb{U}$ are mapped onto the physically allowed domain of $\mathbb{X}$, we found that limiting the domain improved numerical stability in our experiments by avoiding dilute tails in the latent distribution.

We now apply the projection function $h$ to all our datasets with nonzero values of ${\tilde c}_W$\footnote{For simplicity, in this section we only consider variations of ${\tilde c}_W$ and fix $c_{\rm HWB}=0$ throughout.}. It is crucial that $h$ are derived using data at a single point in parameter space (here ${\tilde c}_W=0$) and applied to the data at all values of ${\tilde c}_W$. As ${\tilde c}_W$ is varied, the probability density $p\left(u|{\tilde c}_W\right)$ is deformed. This is modeled as $p_\phi\left(u|{\tilde c}_W\right)$ where the neural network parameters $\phi$ are trained using maximum likelihood estimation evaluated over the simulated training data for all ${\tilde c}_W$, i.e.
\begin{equation}
    \mathbb{V}\left(\phi\right) = \frac{1}{\sum w} ~ \cdot ~ \sum_{{\tilde c}_W,x,w} w \cdot\log p_\phi\left(h\left(x\right)|{\tilde c}_W\right)
\end{equation}
\begin{equation}
    \phi ~\rightarrow~ \rm \mathop{\mathrm{argmax}}_{\phi} ~~ \mathbb{V}\left(\phi\right)
\end{equation}
where $w$ label Monte Carlo event weights, used to account for how integration of probabilities is handled within a particular simulation package\,\cite{Buckley:2011ms,Zyla:2020zbs}, if applicable.

\begin{figure}[htbp!]
    \centering
    \includegraphics[width=0.98\textwidth]{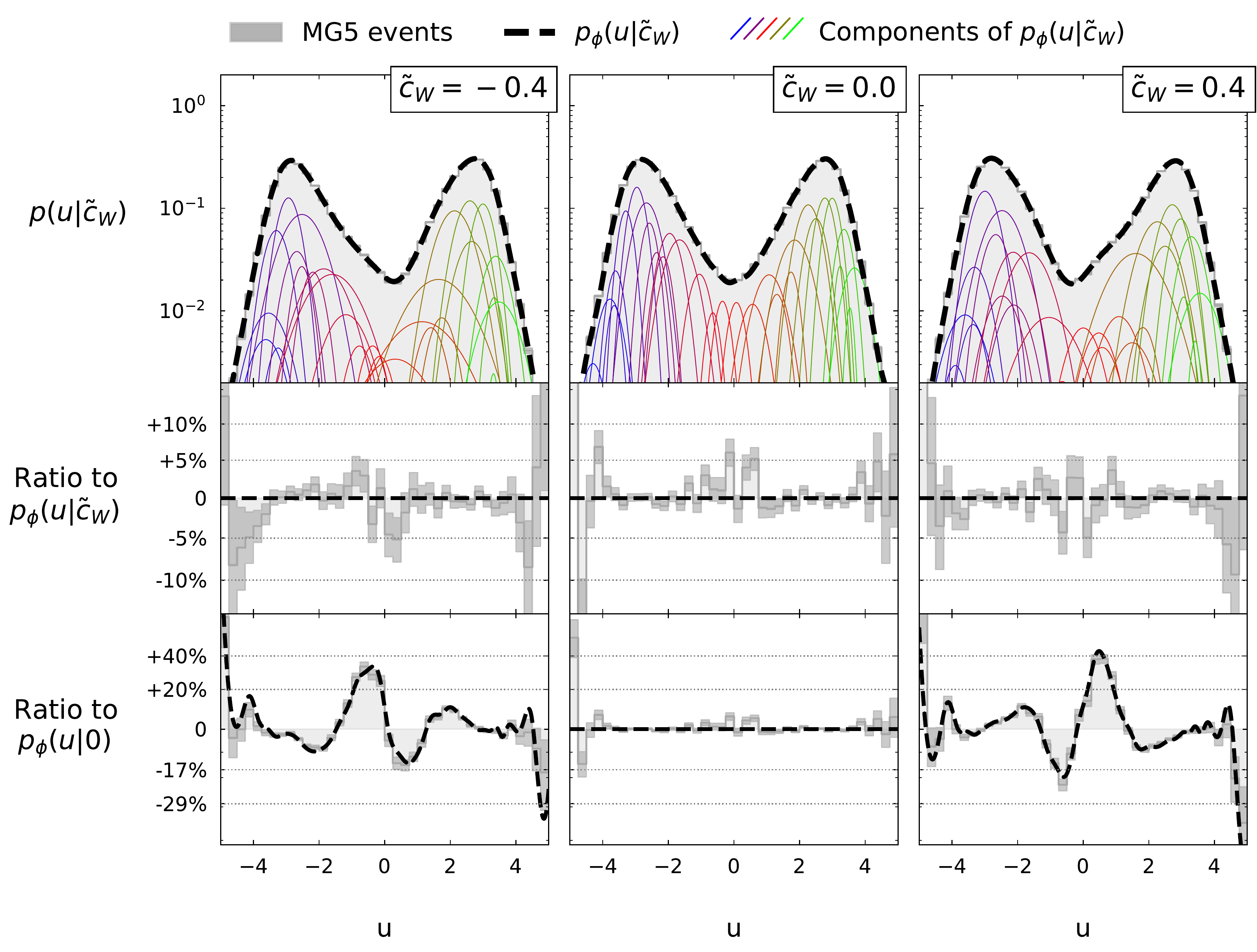}
    \caption{Gaussian mixture model over the latent space (black dashed line) for the one-dimensional example of $x=\Delta\phi\left(j,j\right)$. We show the comparison with \texttt{MG5} events (grey line) when ${\tilde c}_W=-0.4$ (left), ${\tilde c}_W=0$ (middle) and ${\tilde c}_W=0.4$ (right), with $c_{\rm HWB}=0$ throughout. Dark shaded regions show the estimated statistical uncertainty due to the finite number of \texttt{MG5} events. Lower panels show the ratio with respect to $p_\phi\left(u|{\tilde c}_W\right)$ (middle row) and $p_\phi\left(u|0\right)$ (bottom row).}
    \label{fig:proj-fits-1D}
\end{figure}

We train a GMM with $N_G=30$ individual modes to describe the probability density. Figure~\ref{fig:proj-fits-1D} (top row) compares the training data and post-fit model $p_\phi\left(u|{\tilde c}_W\right)$ at values of ${\tilde c}_W=\{-0.4,~0,~0.4\}$. Thin colored lines show the  decomposition into individual Gaussian modes. As ${\tilde c}_W$ is varied, we see that deformations in the spectrum are captured by modulating the amplitudes, positions and widths of the narrow Gaussian modes. 

Figure~\ref{fig:proj-fits-1D} (middle row) shows the ratio between the training data and the model PDF, offset to $0$ so we study the residual difference between the two. This demonstrates that systematic mis-modelling is below $5\%$ except in the sparsely populated tails of the distribution for all three values of ${\tilde c}_W$. The dark shaded band around the data shows the Poisson estimate of the statistical uncertainty. The thickness of this band is comparable with the residual difference between the data and the model, suggesting that this residual is mostly dominated by random fluctuations in the data.

Figure~\ref{fig:proj-fits-1D} (bottom row) shows the ratio between $p_\phi\left(u|{\tilde c}_W\right)$ and $p_\phi\left(u|0\right)$, the model PDF evaluated at ${\tilde c}_W=0$, once again offset to $0$ so we study the residual difference between the two. This quantifies how the shape of the distribution is deformed when translating across ${\tilde c}_W$. Training data are also shown, demonstrating that the model has captured how the spectrum is deformed as ${\tilde c}_W$ is varied.

\section*{Extending to multiple observables}

When modeling $d$ observables on the latent space, we write an auto-regressive probability density
\begin{equation}
p_\phi\left(u|\theta\right) = \prod_{i=1}^{d}p_{\phi, i}\left(u_i|u_{<i},\theta\right)
\end{equation}
where $i$ label observables and $u_{<i}$ is the list of all prior latent observables. The conditional probability density for each $u_i$ is modeled using a GMM parameterized by a neural network with parameters $\phi$ according to
\begin{equation}
    p_{\phi,i}\left(u_i|u_{<i},\theta\right) = \sum_{g=1}^{N_G} f_{\phi, g, i}\left(u_{<i},\theta\right) \cdot \mathcal{N}\left(u_i;~\mu_{\phi, g, i}\left(u_{<i},\theta\right);~\sigma_{\phi, g, i}\left(u_{<i},\theta\right) \right)
\end{equation} where $f_{\phi, g, i}$, $\mu_{\phi, g, i}$ and $\sigma_{\phi, g, i}$ are respectively the amplitude, mean and width of the $g^{\rm th}$ Gaussian for observable index $i$. By including $u_{<i}$ as input to the network, it now captures the dependence on \textit{both} external parameters \textit{and} preceding observables. This means that high-dimensional observable correlations may be described by the model.

\section*{Neural network architecture}

\begin{figure}[h!tbp]
    \centering
\hspace*{1cm}
\includegraphics[width=\textwidth, trim=0 6cm 0 0]{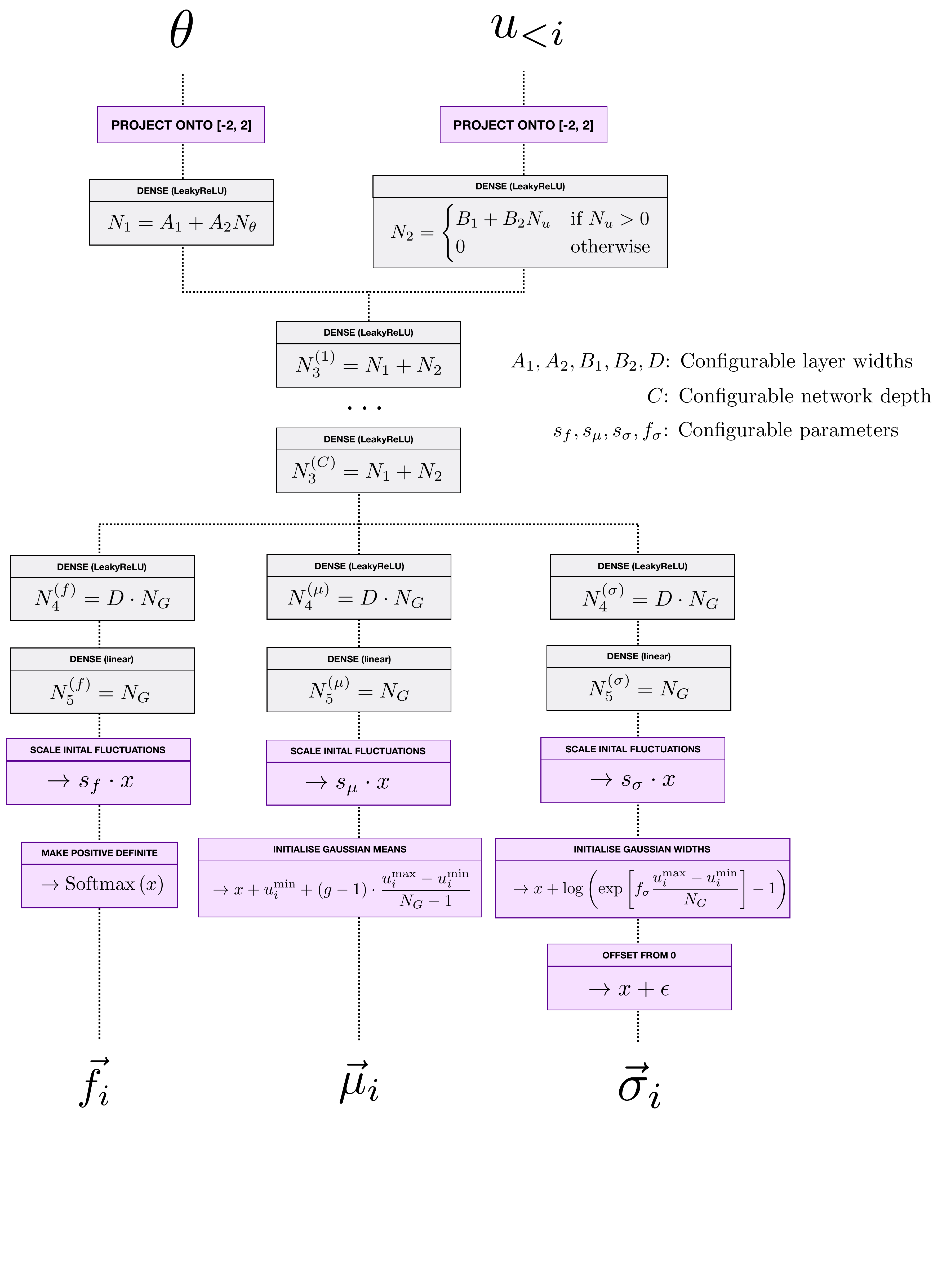}
\hspace*{-1cm}
    \caption{Structure of the neural network implemented for observable $u_i \in \left[u_i^{\rm min},u_i^{\rm max}\right]$. Configurable parameters $\{A_1,A_2,B_1,B_2,D\}$ determine the width of the fully connected Dense layers, which have nodes equal to the $N$ provided, and $C$ determines the number of intermediate Dense layers. Configurable constants $\{s_f,s_\mu,s_\sigma\}$ determine the scale of initial perturbations, while $f_\sigma$ configures the initial Gaussian widths.}
    \label{fig:network-arch}
\end{figure}

Figure~\ref{fig:network-arch} shows a schematic diagram of the neural network architecture used to model the GMM for latent observable $u_i \in \left[u_i^{\rm min},u_i^{\rm max}\right]$. Fully connected layers at depth $l$ are shown in grey and labelled \textit{Dense}, with a number of neurons equal to $N_l$ as specified and an activation function shown in parentheses. These are either \textit{linear}, equivalent to applying no activation function, or \textit{LeakyReLU}\,\cite{leakyrelu} with a negative gradient of $0.2$ defined for input $x$ according to
\begin{equation}
\mathrm{LeakyReLU} \left( x \right) ~=~ 
\cases{
  x & if $x \geq 0$ \cr
  0.2 \cdot x & if $x<0$. \cr
}
\end{equation}

Inputs $\theta$ and $u_{<i}$ of lengths $N_\theta$ and $N_u$ respectively are compressed onto the interval $\left[-2,2\right]$ and fed into initial layers of size $N_1$ and $N_2$. The configurable constants $\{A_1,A_2,B_1,B_2\}$ determine the width of these layers. The outputs are concatenated and fed into a sequence of $C$ layers of width $N_1+N_2$. The constant $C$ determines the ultimate depth of the network. The outputs are then fed into three separate channels, which will separately assign the Gaussian amplitudes $\vec f_i$, means $\vec \mu_i$ and widths $\vec \sigma_i$. In each channel, activations $x$ pass through two further dense layers of size $D \cdot N_G$ and $N_G$, creating three vectors of length $N_G$. These are scaled by factors of $s_f$, $s_\mu$ and $s_\sigma$. These scale factors determine the size of the initial fluctuations around the nominal initial values of $\vec f_i$, $\vec \mu_i$ and $\vec \sigma_i$ which are assigned as follows.

In the $\vec f_i$ channel, activations are passed through a Softmax function to ensure the Gaussian amplitudes are positive definite and sum to unity. If $|s_f| \ll 1$ then all components of $\vec f_i$ are initially approximately equal. In the $\vec \mu_i$ channel, a constant is added to the $g^{\rm th}$ vector component such that the Gaussian modes are initially linearly spaced between $u_i^{\rm min}$ and $u_i^{\rm max}$ subject to fluctuations. In the $\vec \sigma_i$ channel, Gaussian widths are initialized to fluctuate around a value of $f_\sigma$ units of $\frac{u_i^{\rm max}-u_i^{\rm min}}{N_G}$. The configurable constant $f_\sigma$ therefore determines how many standard deviations of overlap exist between the initial Gaussian modes. Finally, a constant of $\epsilon = 10^{-4}$ is added to prevent the evaluation of Gaussian modes with zero width. We note that these transformations impact the gradients of the loss function with respect to the three different channels, leading to different learning rates for the amplitudes, means and widths respectively. This likely impacts the post-fit model, and future optimization may be achieved by controlling the balance of these gradients to preferentially enhance model updates in one channel.

The resulting network contains $\mathcal{O}\left( \left(N_1+N_2\right)^{2C} + \left(N_1 + N_2 + N_G\right)D N_G \right)$ trainable parameters. Model optimization is performed using the \texttt{Adam}\,\cite{adam} algorithm with a learning rate of $\lambda_{\rm lr}$. An adaptive learning rate is used, such that $\lambda_{\rm lr}$ is multiplied by a factor of $\lambda_{\rm lr}^{\rm update~factor}<1$ if the training loss does not improve for $\lambda_{\rm lr}^{\rm patience}$ epochs. This mitigates underfitting when the initial $\lambda_{\rm lr}$ is large. Network biases are initialized to zero and weights are drawn randomly from a uniform distribution over the interval 
$\pm 10/(3\sqrt{N_{\rm in}})$ 
where $N_{\rm in}$ is the number of input neurons. This mitigates vanishing/exploding activations and gradients in the initial state.

\section*{Impact of transforming the likelihood}

The function $h$ performs a monotonic one-dimensional change of variables between $x$ and $u$. The probability density $p_u\left(u\right)$ over the latent space may therefore be transformed into a probability density over the original data space $p_x\left(x\right)$ according to
\begin{equation}
    p_x\left(x\right) ~=~ p_u\left(h\left(x\right)\right) \cdot \left|\frac{d h\left(x\right)}{dx}\right|
\label{Eq::likelihood-transform}
\end{equation}
where $h\left(x\right)$ is evaluated using a piecewise linear function calculated from the training data, and so $\left|\frac{d h\left(x\right)}{dx}\right|$ is a step function over $x$. Whilst it leads to a tractable density over $x$, Equation~\ref{Eq::likelihood-transform} contains no dependence on $\theta$. This means that statistical inference
is equivalent when performed on $\mathbb{U}$ and $\mathbb{X}$. Applying such a transformation is therefore not necessary, and we will always perform inference using observations in the latent representation unless stated otherwise.

We also note that the transformation $h\left(x\right)$ must preserve the total probability contained within a span, i.e.\
\begin{equation}
    \int_{x_1}^{x_2} p_x\left(x\right) {\rm d}x ~=~ \int_{h\left(x_1\right)}^{h\left(x_2\right)} p_u\left(u\right) {\rm d}u
    \label{Eq::likelihood-integral-transform}
\end{equation}
and so we can integrate the probability contained within $[x_1,x_2]$ simply by transforming $x_1$ and $x_2$ and performing the integration over the latent space. However, this integration may only be performed analytically when data are one-dimensional.

We do not perform a rotation when transforming between $x$ and $u$.
This secures three desirable features: it ensures a diagonal Jacobian matrix, it retains an easily understood relationship between each component of $x$ and $u$, and it mitigates potential concerns about loss of generalization\,\cite{whitening-problems}.

\section*{Complexity of likelihood evaluation}

Consider that we wish to model $d$ observables, using $d$ neural networks each containing $L$ hidden layers and $W$ neurons per layer. Assuming that $d \ll W$ and $N_G \ll LW$, the calculation of $p\left(u|\theta\right)$ has a complexity of $\mathcal{O}\left(dLW^2\right)$. However, each of the $d$ conditional probability densities may be computed in parallel, resulting in $\mathcal{O}\left(LW^2\right)$ complexity. This may be further accelerated up to a limit of $\mathcal{O}\left(L\right)$ by using a GPU for efficient matrix multiplication. Since $u_{<i}$ are used as input to the networks for all $i>0$, network outputs must be computed separately for every datapoint except in the case of the first observable $u_0$, for which a single pass through the network can be used to provide the Gaussian parameters needed to evaluate every datapoint.

\section*{Complexity of generative sampling}

We have noted that the density model may be sampled, allowing it to be used as a generative model for event simulation. We achieve this by randomly drawing $u_0^* \sim p_{\phi,0}\left(u_0|\theta\right)$, $u_1^* \sim p_{\phi,1}\left(u_1|u_0^*,\theta\right)$ and so on until a datapoint $u^*$ in $d$ dimensions is constructed. This may be transformed back onto data space using $x^*=h^{-1}\left(u^*\right)$.

Since this process is sequential in the latent observables, they may not be simulated in parallel. As with likelihood evaluation, the complexity of sampling is $\mathcal{O}\left(dLW^2\right)$. This may be accelerated up to a limit of $\mathcal{O}\left(dL\right)$ using a GPU. Since $p_{\phi,0}\left(u_0|\theta\right)$ contains no dependence on other observables, many $u_0^*$ may be sampled using a single evaluation of the network. However, sampling $u_i^*$ for $i>0$ requires the network to be evaluated for every datapoint.

\section*{Modelling of systematic uncertainties}

In this work, we focus on the expressive power of the model and do not consider the impact of systematic uncertainties. However, it is crucial that such uncertainties are accounted for when performing a statistical interpretation on a measured dataset. Here we briefly discuss how this may be done, whilst noting the limitations. We note that cross-section uncertainties may be trivially accounted for, since they do not impact the distribution of events throughout phase space.

We may separate modelling uncertainties into three categories. The first category are uncertainties associated with the simulation of training data which are parameterizable in terms of a nuisance parameter $\theta_{\rm NP}$. These may be accounted for either by including $\theta_{\rm NP}$ within the vector $\theta$ input to the network, or by training a separate model 
$r\left(u,\theta_{\rm NP}\right) = p\left(u|\theta_{\rm NP}\right) / p\left(u|\theta_{\rm NP}^{\rm ref}\right)$ 
for some fixed reference $\theta_\mathrm{NP}^\mathrm{ref}$ and writing
\begin{equation}
    p\left(u|\theta_{\rm NP}\right) ~=~ p\left(u|\theta_{\rm NP}^{\rm ref}\right) \cdot r\left(u,\theta_{\rm NP}\right) 
    ~~.
\end{equation}

The second category are non-parameterizable uncertainties associated with the simulation of training data. In high energy physics, these may account for poorly understood differences between the simulated data and control measurements. In a binned one-dimensional analysis, they may be mitigated by performing auxiliary observations which are uncorrelated with the observable being modelled and ``transferring'' the data-driven constraint on a bin-by-bin basis. Residual uncertainties may then be parameterized according to systematic variations of this transfer procedure. It is challenging to extend such techniques to our model because we must cover possible mismodelling of the high-dimensional observable correlations.

The third category are uncertainties associated with the density model. These biases are caused by the inductive bias of the model as well as under- or over-fitting. Over-fitting may be mitigated using techniques such as regularization, dropout and early stopping, and by limiting model complexity. Under-fitting may be studied by sampling the density model for all simulated $\theta$ and showing that the marginal projections are compatible with the simulated data. Quantifying and parameterizing the remaining mismodelling is once again challenging, and we leave this for future work.

We consider overcoming these challenges to be one of the main hurdles facing the use of high-dimensional density models in high energy physics.

\section*{Model optimization}

A strength of the proposed method is that there are many ways in which modelling may be improved if under-fitting is observed. These strategies include:
\begin{enumerate}
    \item Increase the model capacity by using more complicated networks or larger $N_G$.
    \item Tune the parameters $s_f$, $s_\mu$ and $s_\sigma$, which modulate the size of the initial state perturbations of the Gaussian amplitudes, positions and widths as described in Figure~\ref{fig:network-arch}, to balance the stability of the initial model with the size of perturbations which provide gradients for the learning process. 
    \item Tune $f_\sigma$ to configure the initial width of the Gaussian modes. Whilst narrow modes tend to describe local features of the data, fulfilling the objectives of our model design, training data do not provide significant learning potential for Gaussian modes several standard deviations away. We find that successful training occurs when the value of $f_\sigma$ balances these effects.
    \item Tune the hyperparameter $f$ or the functional form of ${\tilde q}_u$ to create a latent distribution which is well described by a mixture of narrow Gaussians.
    \item Alter the ordering of the observables, since $p\left(B|A\right)$ may be more easily described than $p\left(A|B\right)$ for two latent observables $A$ and $B$.
    \item Alter the training procedure to improve convergence towards likelihood maxima.
    \item Rotate observables onto the eigenvectors of their covariance, reducing strong correlations in the data.
\end{enumerate}
These opportunities for tuning improve the chance of finding a model which captures the salient features of the dataset provided.

\section{EW Zjj with 12 observables and no external parameter dependence}
\label{sec:VBFZ-11obs}

In this section we create a density model to describe 12 observables with no external parameter dependence. This demonstrates that the method can learn a joint probability density over a high-dimensional dataset of physically realistic observables. Table~\ref{tab:VBFZ-0D-whitening-constants} shows the observable ordering as well as the $f$-values used to configure the projection onto the latent space.
\begin{table}[htbp]
    \caption{Indices in which observables are ordered when constructing a density model describing EW $Zjj$ data with 12 observables and no external parameter dependence. The $f$ values used to project continuous real-valued observables onto the latent space are shown. Indices start from $0$.}
    \label{tab:VBFZ-0D-whitening-constants}
\begin{indented}
\item[]\begin{tabular}{@{}rll rll rll rll }
\br
& \multicolumn{7}{c}{Observable order: name [projection constant $f$]} & \\
\mr
$0$:& $m_{\rm jj}$ & [$f=0.2$]  &  $1$:& $p_{\rm T}^{\rm jj}$ & [$f=0.2$]  &  $2$:& $|y^{\rm jj}|$ & [$f=0.2$]  \\

         $3$:& $\Delta\phi\left(j,j\right)$ & [$f=0.8$]  &  $4$:& $\Delta y\left(j,j\right)$ & [$f=0.8$]  &  $5$:& $p_{\rm T}^{\rm j1}$ & [$f=0.2$]  \\

         $6$:& $p_{\rm T}^{\rm j2}$ & [$f=0.2$]  &  $7$:& $N_{\rm gapjet}$ &  &  $8$:& $N_{\rm jet}$ &  \\

         $9$:& $m_{\rm ll}$ & [$f=0.8$]  &  $10$:& $p_{\rm T}^{\rm ll}$ & [$f=0.2$]  &  $11$:& $|y^{\rm ll}|$ & [$f=0.8$] \\
\br
\end{tabular}
\end{indented}
\end{table}

We include the two discrete observables $N_{\rm gapjet}$ and $N_{\rm jet}$ in the model. This demonstrates that there are no barriers to modelling continuous and discrete observables at the same time. A discrete observable taking integer values on the inclusive interval $[u_i^{\rm min},u_i^{\rm max}]$ is modelled using a neural network which outputs a categorical probability distribution of length $N_p = 1+u_i^{\rm max}-u_i^{\rm min}$. Inputs $\theta$ and $u_{<i}$ are projected onto the interval $[-2,2]$ and passed through dense layers of size $N_1$ and $N_2$ respectively. These are followed by two fully connected layers of size $300$ and $200$, and an output layer of size $N_p$. All intermediate layers use a LeakyReLU activation function with a negative gradient of $0.2$. The output layer uses a SoftMax activation function to ensure that outputs represent a normalized multinomial probability distribution. The network is trained using a cross entropy loss function and the same training scheme as used to model continuous observables.

Table~\ref{tab:VBFZ-0D-model-constants} shows the constants used to configure the remaining neural networks and their training. The networks contain between $27{\rm k}$ and $304{\rm k}$ trainable parameters. This reflects a degree of over-parameterization of the model, since the number of parameters is the same order of magnitude as the number of training samples. We note that any resultant over-training is mitigated by the use of a GMM which naturally smooths each conditional PDF in the auto-regressive chain. Each network is initially trained for up to 400 epochs, stopping early if the loss function does not improve over a period of 12 epochs. We observe that $\mathcal{O}\left(10^{-4}\right)$ relative updates to the log-likelihood are important, since they may lead to $\%$-level improvements in the description of the tails. Training should therefore not be halted until a true plateau in the loss function is obtained.
\begin{table}[htbp]
    \caption{Constants used to construct and train a density model describing EW $Zjj$ data with 12 observables and no external parameter dependence.}
    \label{tab:VBFZ-0D-model-constants}
\begin{indented}
\item[]\begin{tabular}{@{}ccccc}
\br
         $N_G=20$ & $A_1=200$ & $A_2=0$ & $B_1=200$ & $B_2=50$ \\

         $C=3$ & $D=3$ & $s_f=0.01$ & $s_\mu=0.01$ & $s_\sigma=0.01$ \\

        $f_\sigma=0.5$ & batch size = 1k & $\lambda_{\rm lr}=0.001$  &  $\lambda_{\rm lr}^{\rm update~factor}=0.5$  &  $\lambda_{\rm lr}^{\rm patience}=3$ \\
\br
    \end{tabular}
\end{indented}
\end{table}

The model is trained using the $640{\mathrm k}$ selected \texttt{MG5} events generated assuming the SM hypothesis. To evaluate its performance, we randomly sample $4{\mathrm M}$ datapoints from the model and compare the 1D and 2D marginal distributions with those of the training data. This large number is chosen to reduce fluctuations due to sampling variance.

\begin{figure}[htbp]
    \centering
    \includegraphics[width=0.99\textwidth]{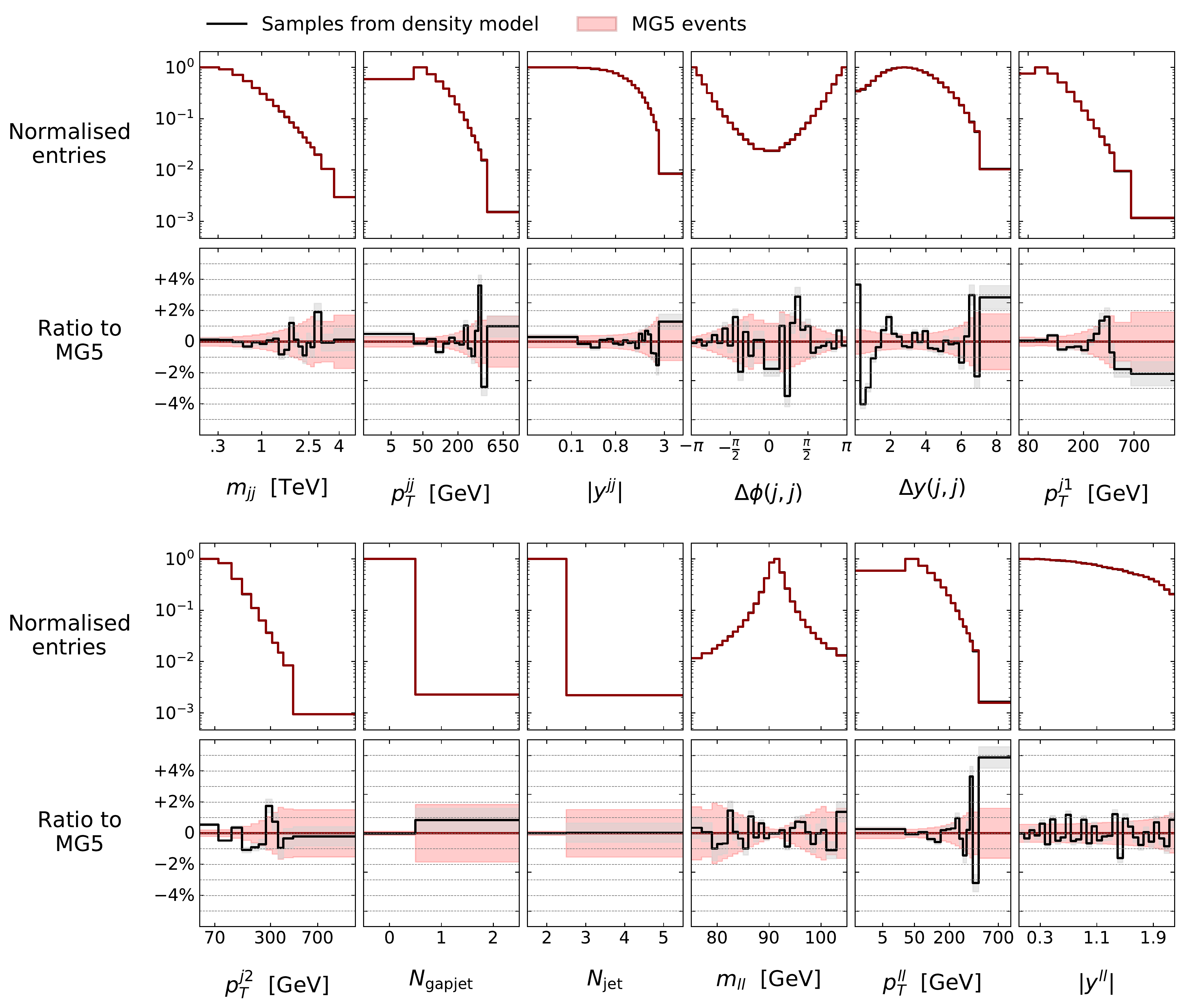} \\
    \caption{1D marginal distributions comparing events simulated with MG5 (red) with those sampled from a GMM trained on a latent space (black) with no external parameter dependence. Note that the two spectra are not statistically independent, since the density model was trained using the \texttt{MG5} events.}
    \label{fig:VBFZ-11obs-1D-dists}
\end{figure}

Figure~\ref{fig:VBFZ-11obs-1D-dists} presents the 1D marginal distributions. For each observable, an upper panel presents the absolute spectrum in units normalized such that the highest bin takes a value of $1$, and a lower panel shows a ratio taken with respect to the \texttt{MG5} events. \texttt{MG5} events are shown in red and compared with events sampled from the density model, shown in black. Shaded areas present Poisson estimates of the statistical uncertainty arising from finite sample size. We observe that all spectra are well described within a systematic precision of $\pm 5~\%$, with many spectra achieving precision similar to the statistical variance of the training data. We note that fewer bins than the expected $\mathcal{O}\left(32~\%\right)$ lie outside of the uncertainty bands, indicating that the model may be over-trained. Since this work is intended as a proof-of-principle for the method, we make no further attempt to mitigate over-training, whilst noting that this will be important for future applications. 

\begin{figure}[htbp]
\centering
    \includegraphics[angle=90,width=\textwidth]{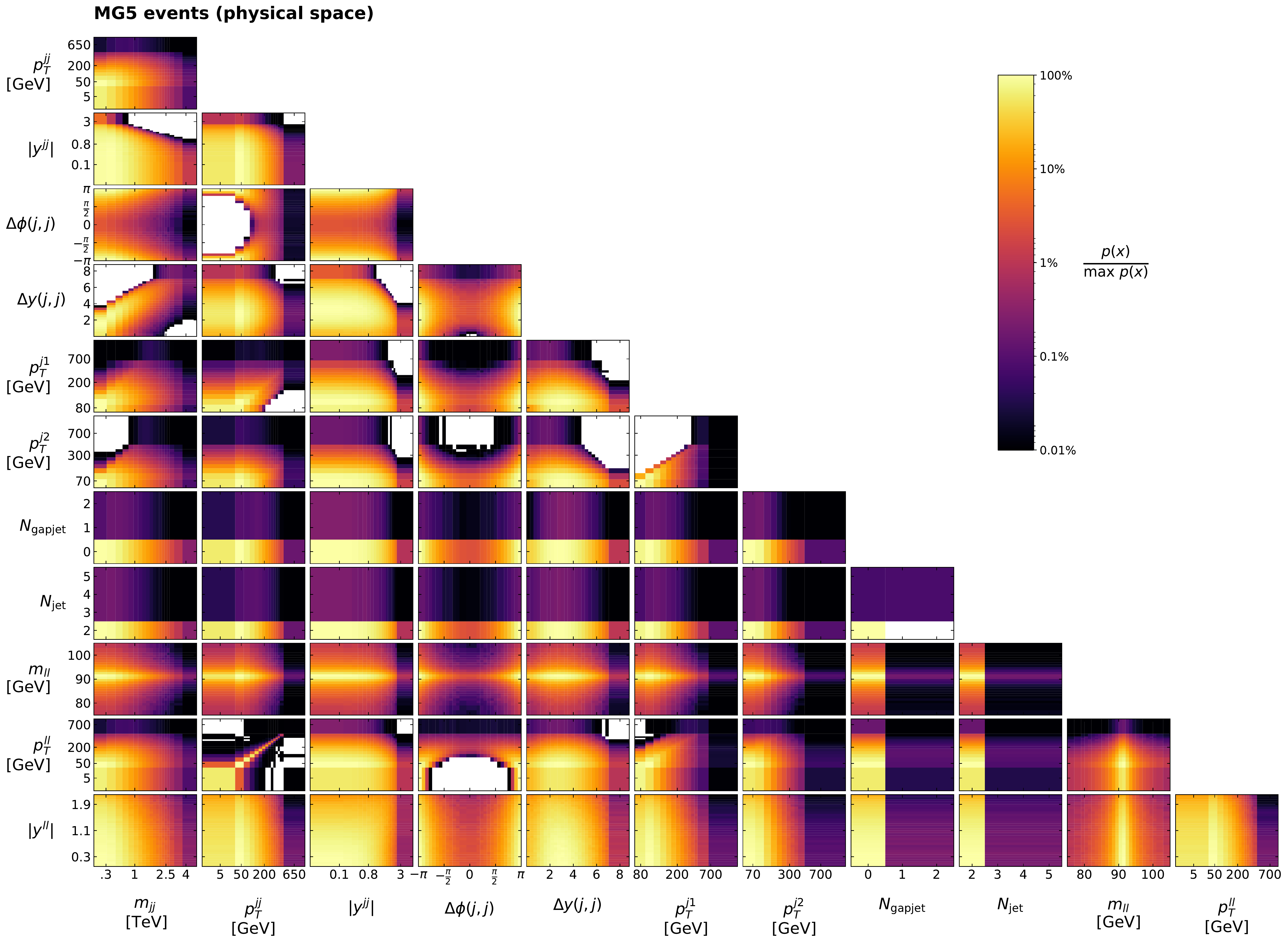} \\
\caption{2D marginal distributions of events simulated with MG5 at the SM hypothesis.}
    \label{fig:VBFZ-11obs-2D-MG}
\end{figure}

\begin{figure}[htbp]
    \centering
    \includegraphics[angle=90,width=\textwidth]{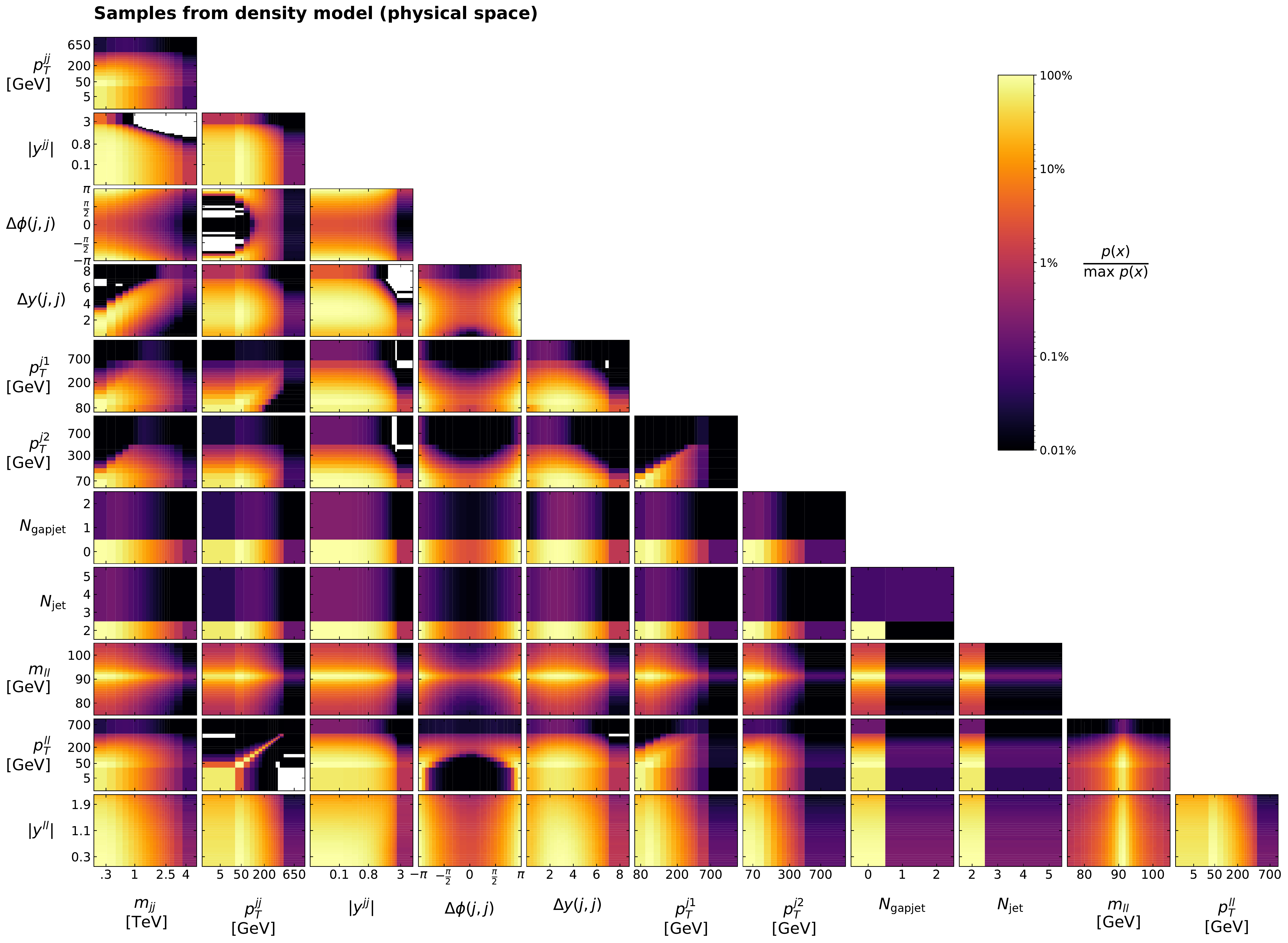} \\
    \caption{2D marginal distributions of events sampled from a GMM trained on a latent space with no external parameter dependence, assuming the SM hypothesis.}
    \label{fig:VBFZ-11obs-2D-GMM}
\end{figure}

Figure~\ref{fig:VBFZ-11obs-2D-MG} presents the 2D marginal distributions for all pairs of observables as measured using the \texttt{MG5} events. This demonstrates that complex correlations exist between all observables. Figure~\ref{fig:VBFZ-11obs-2D-GMM} presents the 2D marginal distributions using the samples from the density model. Comparing Figures~\ref{fig:VBFZ-11obs-2D-MG} and \ref{fig:VBFZ-11obs-2D-GMM} shows that the model has captured the high-dimensional correlations between all pairs of observables. Bins are coloured white if no entries exist, and black if a small number of entries are observed. We note that several fully-white regions of Figure~\ref{fig:VBFZ-11obs-2D-MG} are black in Figure~\ref{fig:VBFZ-11obs-2D-GMM}, suggesting that the density model may predict a small non-zero probability in regions of phase space which are unpopulated when simulating from-first-principles, as is the case with \texttt{MG5}.

If the modelled density in such regions is sufficiently small, we expect that this artifact should have minimal impact on inference tasks. This is because any overflow of density into physically-disallowed regions of phase space will mainly cause a small under-estimate of the normalization in physically-allowed regions, where all observed events must necessarily exist. Furthermore, this normalization shift may cancel when considering likelihood ratios. A greater problem may occur when using the density model for event sampling, since events may be generated in the physically-disallowed regions. Whilst not solving this problem at this time, we foresee potential for mitigation using two methods:
\begin{enumerate}
    \item Use transformed observables which enforce easily-parameterized boundaries. For example, modelling the pair of observables $\{p_T^{\rm j1},p_T^{\rm j2}\}$ risks predicting a non-zero density in the unphysical region $p_T^{\rm j2} > p_T^{\rm j1}$. Instead we
    can model $\{p_T^{\rm j1}{'}, p_T^{\rm j2}\}$ where $p_T^{\rm j1}{'} = p_T^{\rm j1} - p_T^{\rm j2}$ is required to satisfy $p_T^{\rm j1}{'}\geq0$, preventing such unphysical behaviour.
    A drawback is that we cannot enforce the original boundary limits of $p_T^{\rm j1}$, because these must now be defined relative to the value of $p_T^{\rm j2}$. Furthermore, most physical boundary conditions may not be easily enforced by such a transformation, either because they are too complicated or because the user is not aware of them.
    \item In high energy physics, one can model the components of object four-vectors and reconstruct observables accordingly. This naturally imposes many physical constraints, although not all, and once again we cannot enforce simple boundary conditions for high-level observables.
\end{enumerate}

\begin{figure}[htbp]
\centering
    \includegraphics[angle=90,width=.95\textwidth]{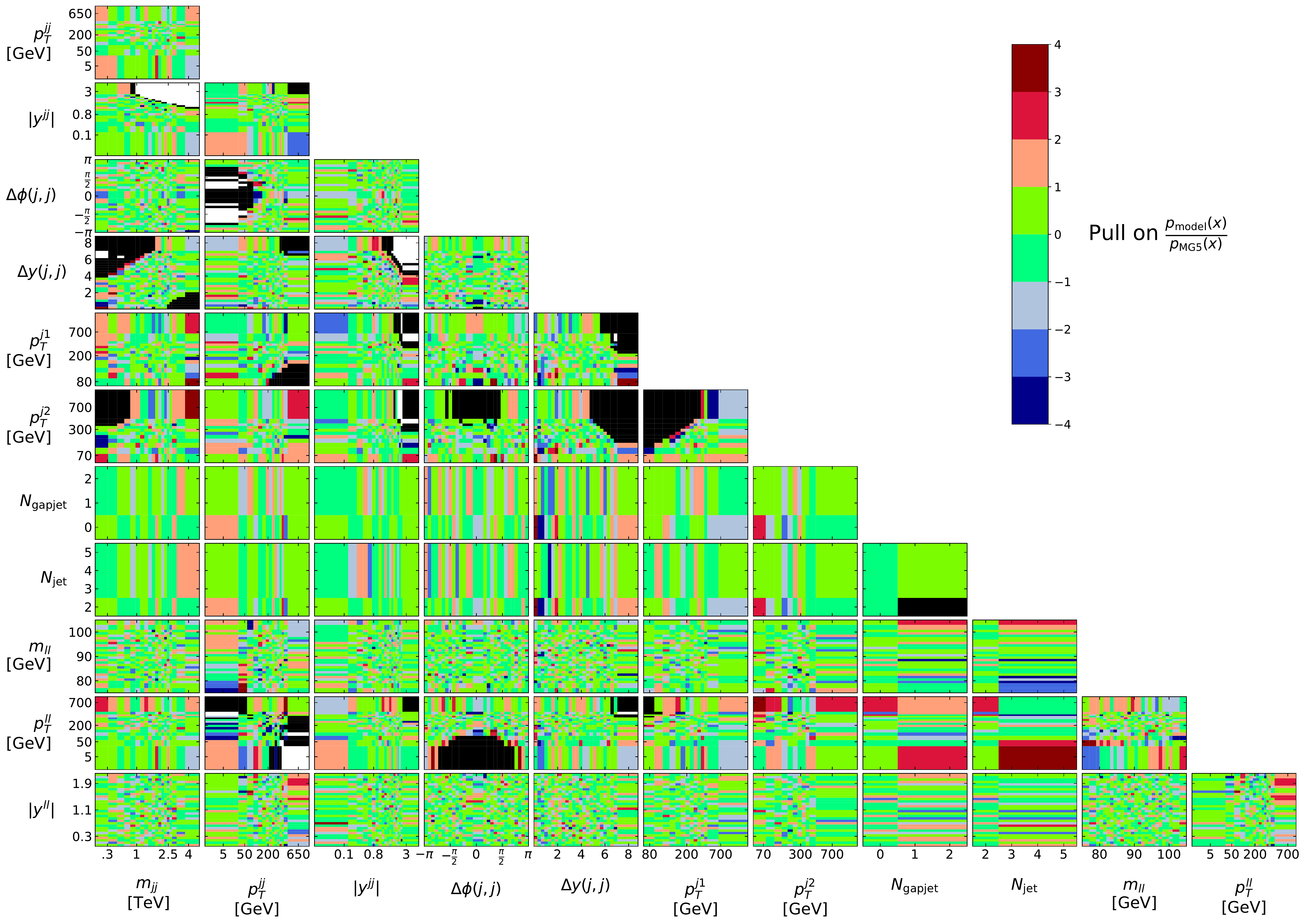} \\
    \caption{Pull on the ratio between the 2D marginal distributions comparing events simulated with MG5 (denominator) with those sampled from a GMM trained on a latent space (numerator). The model has no external parameter dependence.}
    \label{fig:VBFZ-11obs-2D-ratios}
\end{figure}

With these caveats, Figures~\ref{fig:VBFZ-11obs-2D-MG} and \ref{fig:VBFZ-11obs-2D-GMM} demonstrate that the 2D projections of events sampled using density model are qualitatively very similar to the ground truth events throughout most of the space. The comparison is quantified in Figure~\ref{fig:VBFZ-11obs-2D-ratios}. This shows the pull on the ratio of these histograms, defined as
\begin{equation}
    \mathrm{Pull~on~} \frac{p_{\rm model}}{p_{\rm MG5}} ~=~ \frac{\frac{p_{\rm model}~-~p_{\rm MG5}}{p_{\rm MG5}}}{\Delta\left( \frac{p_{\rm model}}{p_{\rm MG5}}\right)}
\end{equation}
where $p_{\rm model}$ and $p_{\rm MG5}$ are the densities estimated using events sampled from the density model and \texttt{MG5} respectively, and $\Delta\left(\frac{p_{\rm model}}{p_{\rm MG5}}\right)$ represents the estimated statistical uncertainty on the ratio between them. This dominated by the estimated statistical uncertainty on $p_{\rm MG5}$. The pull can be interpreted as ``the number of standard deviations by which the ratio differs from unity.'' It therefore shows the sign and statistical significance of the difference between the two distributions. 

If the density model represents an unbiased fit to the \texttt{MG5} events then we expect $O\left(68\%\right)$ of bins to fall inside the interval $[-1,+1]$. Due to random fluctuations in the event sampling, we still expect $O\left(32\%\right)$ to fall outside of this interval by chance, even when the density model is equal to the ground truth. Extending this idea, we expect $O\left(5\%\right)$ of bins to fall outside the interval $[-2,+2]$ and $O\left(0.3\%\right)$ outside $[-3,+3]$ due to random fluctuations. If the density model is over-trained, we expect to observe an excess of bins with small pulls. Where mis-modeling occurs, we expect to observe a systematic trend of large pulls.

This allows us to study the agreement between the density model and training events in the following way. All bins with pulls less than $1$ in magnitude are shown in green in Figure~\ref{fig:VBFZ-11obs-2D-ratios}. These are bins where the agreement between the density model and training data is better than the estimated statistical uncertainty. Bins with pulls above $+1$ (below $-1$) are shown in increasingly dark shades of red (blue). Since we expect that $O\left(32\%\right)$ of bins will be colored red or blue, the presence of these bins does not indicate mis-modeling. Instead we search for the following signatures:
\begin{itemize}
    \item Adjacent dark red or blue bins indicate that the difference between the density model and training data is unlikely to occur by chance. These regions are likely mis-modeled.
    \item Multiple adjacent bins which are all colored red or blue suggest an effect of systematic mis-modeling rather than statistical fluctuation.
    \item More bins shaded in red or blue than expected, indicating that more bins than expected exceed the statistical variance due to sampling, suggesting that mis-modeling is present in some of these regions.
\end{itemize}

Since most bins in Figure~\ref{fig:VBFZ-11obs-2D-ratios} are colored green or light red/blue, we observe that most of the space is well-described within $\pm 2$ standard deviations. This indicates that, in general, the model is able to describe the high-dimensional distribution of the data at a level comparable with the statistical precision of the training data.

Some red or blue bands are observed, for example (i) in the steeply falling tail of the $\Delta y\left(j,j\right)$ distribution when projected along with $m_{jj}$ and $|y^{jj}|$, and (ii) in the region $p_{\rm T}^{j1} \approx p_{\rm T}^{j2}$ in the projection of the two. This suggests some systematic mis-modeling in these regions, and scope for tuning using the optimization methods suggested in section~\ref{sec:method}.

White regions indicate that no density is present, whilst black regions indicate that events are present when sampling the density model but not \texttt{MG5}, repeating the observations discussed above. Table~\ref{tab:VBFZ-0D-model-pull_freqs} summarizes the total frequency with which pulls are observed in the different color bins.

\begin{table}[htbp]
    \caption{Frequencies of pulls observed in Figure~\ref{fig:VBFZ-11obs-2D-ratios}, ignoring black and white bins which contain zero sampled events.}
    \label{tab:VBFZ-0D-model-pull_freqs}
\begin{indented}
\item[]\begin{tabular}{@{}r@{\hskip 20pt}r@{.}lr@{.}l}
\br
 Pull range &  \multicolumn{2}{c}{Observed frequency}  &  \multicolumn{2}{c}{Expected frequency}  \\
\br
  Below $-4$    &  \qquad$\ll0$&$003~\%$  &  \quad$\mathcal{O}(0$&$003~\%)$ \\
  $-4$ to $-3$  &  \qquad$0$&$58~\%$      &  \quad$\mathcal{O}(0$&$13~\%)$  \\
  $-3$ to $-2$  &  \qquad$3$&$4~\%$       &  \quad$\mathcal{O}(2$&$1~\%)$   \\
  $-2$ to $-1$  &  \qquad$13$&$4~\%$      &  \quad$\mathcal{O}(13$&$6~\%)$  \\
  $-1$ to $0$   &  \qquad$31$&$6~\%$      &  \quad$\mathcal{O}(34$&$1~\%)$  \\
  $0$ to $+1$    &  \qquad$35$&$0~\%$      &  \quad$\mathcal{O}(34$&$1~\%)$   \\
  $+1$ to $+2$    &  \qquad$13$&$7~\%$      &  \quad$\mathcal{O}(13$&$6~\%)$   \\
  $+2$ to $+3$    &  \qquad$1$&$8~\%$       &  \quad$\mathcal{O}(2$&$1~\%)$    \\
  $+3$ to $+4$    &  \qquad$0$&$17~\%$      &  \quad$\mathcal{O}(0$&$13~\%)$   \\
  Above $+4$     &  \qquad$\ll0$&$003~\%$  &  \quad$\mathcal{O}(0$&$003~\%)$  \\
\br         
    \end{tabular}
\end{indented}
\end{table}

\section{EW Zjj with 4 observables and 2 external parameters}
\label{sec:VBFZ-4obs-2param}

We now train a model which captures the dependence of EW $Zjj$ data on the external parameters $\vec c = \{c_{\rm HWB},~{\tilde c}_W\}$. Such a model may be used to perform maximum likelihood estimation or derive exclusion limits on the space of $\vec c$ based on an observed dataset\footnote{We emphasize that detector effects have not been applied to our training data, but would be for such an analysis.}.

In this case, two SMEFT coefficients would be profiled with all others assumed to be $0$. This is consistent with experimental analyses in the Higgs and electroweak sectors, in which only one or two parameters are usually profiled at a time. In general, it is not possible to constrain many more parameters. This is because we must simulate training data at regular intervals in all directions of $\vec c$. The number of required simulations therefore grows exponentially with the number of parameters profiled, which quickly becomes computationally intractable\footnote{We note that global fits of SMEFT parameters are possible when using binned measurements \cite{Ellis_2018,Ellis_2021,brivio2021models}. This is because the prediction for a given $\vec c$ may be decomposed into a parametric relationship between a number of pure-SM, pure-SMEFT and interference terms. In this case, since the number of unique terms rises slower than exponentially with the number of parameters, all parameters which impact EW $Zjj$ events may be profiled together. However, for general new physics models where no such parameterisation exists, the number of parameters profiled will be limited by the curse-of-dimensionality. Exploiting this special case is not possible using our method because we cannot express negative event densities which may arise in the interference term.}.

We note that the external parameters also impact the \textit{rate} $\sigma_{\rm fid}\left(\vec c\right)$ at which signal is expected to be produced within the observable phase space. When performing an experiment with a fixed exposure (rather than a fixed number of events), we expect to observe events at a point $x$ in phase space at a rate of
\begin{equation}
    \frac{{\rm d}\sigma\left(x|\vec c\right)}{{\rm d}x} ~=~ \sigma_{\rm fid}\left(\vec c\right) ~\cdot ~ p\left(x|\vec c\right) ~~.
\end{equation}
In this work we consider the modeling of $p\left(x|\vec c\right)$. We note that $\sigma_{\rm fid}\left(\vec c\right)$ may typically be modelled using a simple feed-forward neural network, allowing the event rate to be used as a discriminating observable if desired.

We also note that we are not modeling any backgrounds to the EW $Zjj$ process. This is because we wish to test our ability to model multi-dimensional data with a non-trivial parameter dependence. This is best achieved by isolating the signal component, since in general background processes will not depend on the same parameters. However, we note that background modeling must be considered when performing parameter inference using detector-level data, and in particular a large irreducible background from non-electroweak $Zjj$ production would exist in a `real-world' EW $Zjj$ analysis. For such an analysis, a statistical model combining individual components $p_{\rm sig}\left(x|\vec c\right)$ and $p_{\rm bkg}\left(x\right)$ with expected cross-sections $\sigma_{\rm sig}\left(\vec c\right)$ and $\sigma_{\rm bkg}$ may be constructed as
\begin{equation}
    p\left(x|\vec c\right) ~=~ \frac{\sigma_{\rm sig}\left(\vec c\right) \cdot p_{\rm sig}\left(x|\vec c\right) ~+~ \sigma_{\rm bkg} \cdot p_{\rm bkg}\left(x\right)}{\sigma_{\rm sig}\left(\vec c\right) ~+~ \sigma_{\rm bkg}}
\end{equation}
assuming that interference is either small or absorbed into the background model.

For simplicity we select four observables to model, in the sequential order $p_T^{ll}$, $p_T^{j1}$, $m_{jj}$ and finally $\Delta\phi\left(j,j\right)$, excluding the other eight from consideration. All four observables are expected to depend on the external parameters, and we aim to capture this dependence within our model.

The projection onto the latent space is performed using the same $f$-values as presented in Table~\ref{tab:VBFZ-0D-whitening-constants} and used in the previous section. Table~\ref{tab:VBFZ-2D-model-constants} presents the constants used to configure the neural networks which contain $18{\rm k}-85{\rm k}$ trainable parameters. Compared with those in Table~\ref{tab:VBFZ-0D-model-constants}, we note that larger values of $s_f$, $s_\mu$ and $s_\sigma$ are used. This initializes the model such that external parameter variations deform the kinematic spectra, and so impact the log-likelihood, significantly enough that we find an improved parameter dependence to be learned during training. However, we note that large values may excessively enhance fluctuations and lead to an unstable initial state, and the final constants are chosen to balance these effects. The constant $f_\sigma$ is tuned to ensure that the initial Gaussian width is not much larger than the scale of latent space features which are deformed by parameter variations.

\begin{table}[htbp]
    \caption{Constants used to construct and train a density model describing EW $Zjj$ data with 4 observables and 2 external parameters.}
    \label{tab:VBFZ-2D-model-constants}
\begin{indented}
\item[]\begin{tabular}{@{}ccccc}
\br         
$N_G=30$ & $A_1=50$ & $A_2=0$ & $B_1=50$ & $B_2=20$ \\

         $C=2$ & $D=3$ & $s_f=0.125$ & $s_\mu=0.125$ & $s_\sigma=0.125$ \\

        $f_\sigma=0.25$ & batch size = 5k & $\lambda_{\rm lr}=0.001$  &  $\lambda_{\rm lr}^{\rm update~factor}=0.5$  &  $\lambda_{\rm lr}^{\rm patience}=3$ \\
\br         
    \end{tabular}
\end{indented}
\end{table}

\begin{figure}[p!]
    \centering
    \includegraphics[width=0.9\textwidth]{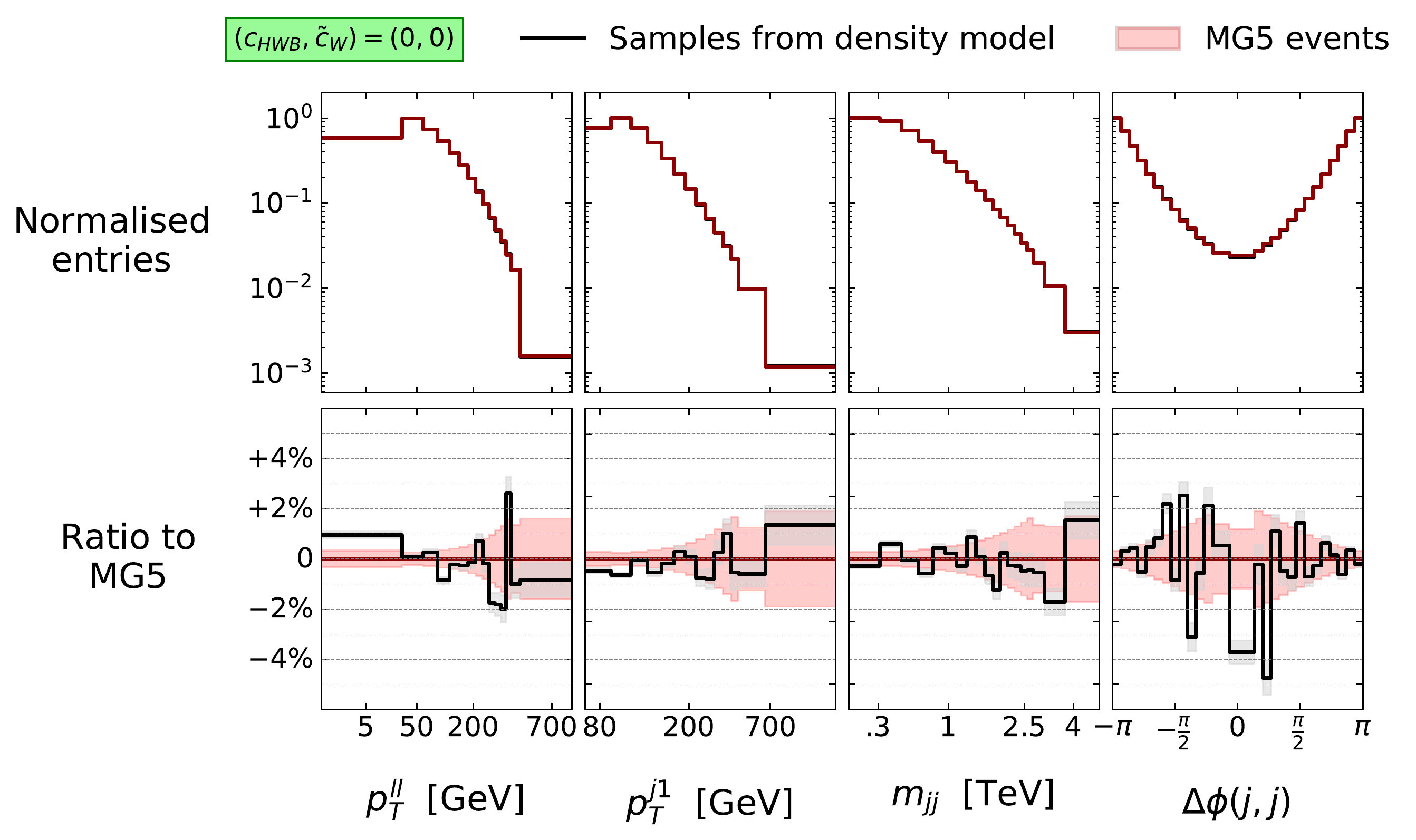}
    \caption{Marginal distributions of events sampled using the density model (black) compared with those generated using \texttt{MG5} (red) for a value of $\left(c_{\rm HWB},{\tilde c}_W\right)=\left(0,0\right)$. Shaded areas show sampling uncertainties. Note that the two spectra are not statistically independent, since the density model was trained using the \texttt{MG5} events.}
    \label{fig:VBFZ-4obs-1D-dist}
    \vspace{1.0cm}
    \includegraphics[width=0.9\textwidth]{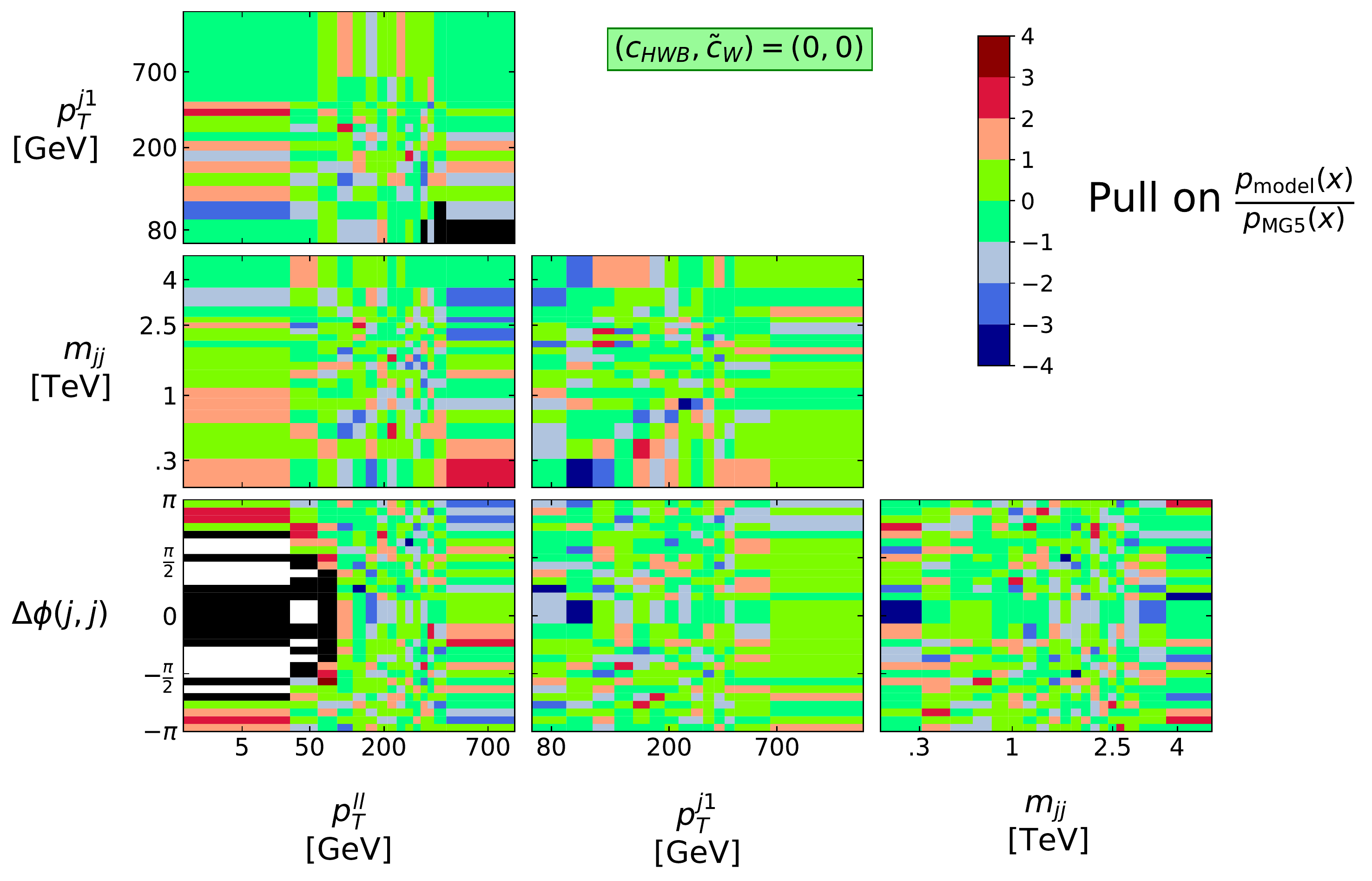}
    \caption{Pull on the ratio between the 2D marginal distributions comparing events simulated with MG5 (denominator) with those sampled from a GMM trained on a latent space (numerator), both assuming the SM hypothesis of $\left(c_{\rm HWB},{\tilde c}_W\right)=\left(0,0\right)$. The model accepts $c_{\rm HWB}$ and ${\tilde c}_W$ as input parameters.}
    \label{fig:VBFZ-4obs-2D-dist}
\end{figure}

Each neural network is trained for up to 200 epochs, stopping early if the log-likelihood does not improve by an amount greater than $10^{-10}$ over a period of 15 epochs. Figure~\ref{fig:VBFZ-4obs-1D-dist} shows the 1D marginal distributions evaluated at the SM hypothesis of $\vec c=\left(0,0\right)$, obtained by sampling $4M$ events from the density model. Figure~\ref{fig:VBFZ-4obs-2D-dist} shows the corresponding pulls on the 2D marginal spectra. Replicating the results of the previous section, these demonstrate that the model describes the 1D distributions to within $\pm 5\%$ at this point in parameter space, and without significant pulls in the 2D projections.

To investigate whether the parameter dependence has been learned, we scan across all hypotheses in the $\vec c$-plane and study the \textit{ratio} of the 1D marginal distributions when compared with the SM. To reduce sampling variance when studying the density model, we form this ratio using importance sampling. We first sample $100k$ events from the model assuming the SM hypothesis. We then use the density model to evaluate the probability density of every datapoint under both the SM and $\vec c$ hypotheses, labelled $p_\mathrm{SM}$ and $p_c$ respectively. The distribution under the $\vec c$ hypothesis is then obtained by assigning a weight of $\frac{p_c}{p_\mathrm{SM}}$ to every datapoint. This approach assumes that the probability distribution under the SM hypothesis fully spans the support of that of the $\vec c$ hypothesis. The result is that the distributions obtained under the SM and $\vec c$ hypotheses have strongly correlated statistical fluctuations. These largely cancel when we take the ratio, which can be estimated using fewer samples than if the hypotheses were sampled independently.

\begin{figure}[p!]
    \includegraphics[width=\textwidth]{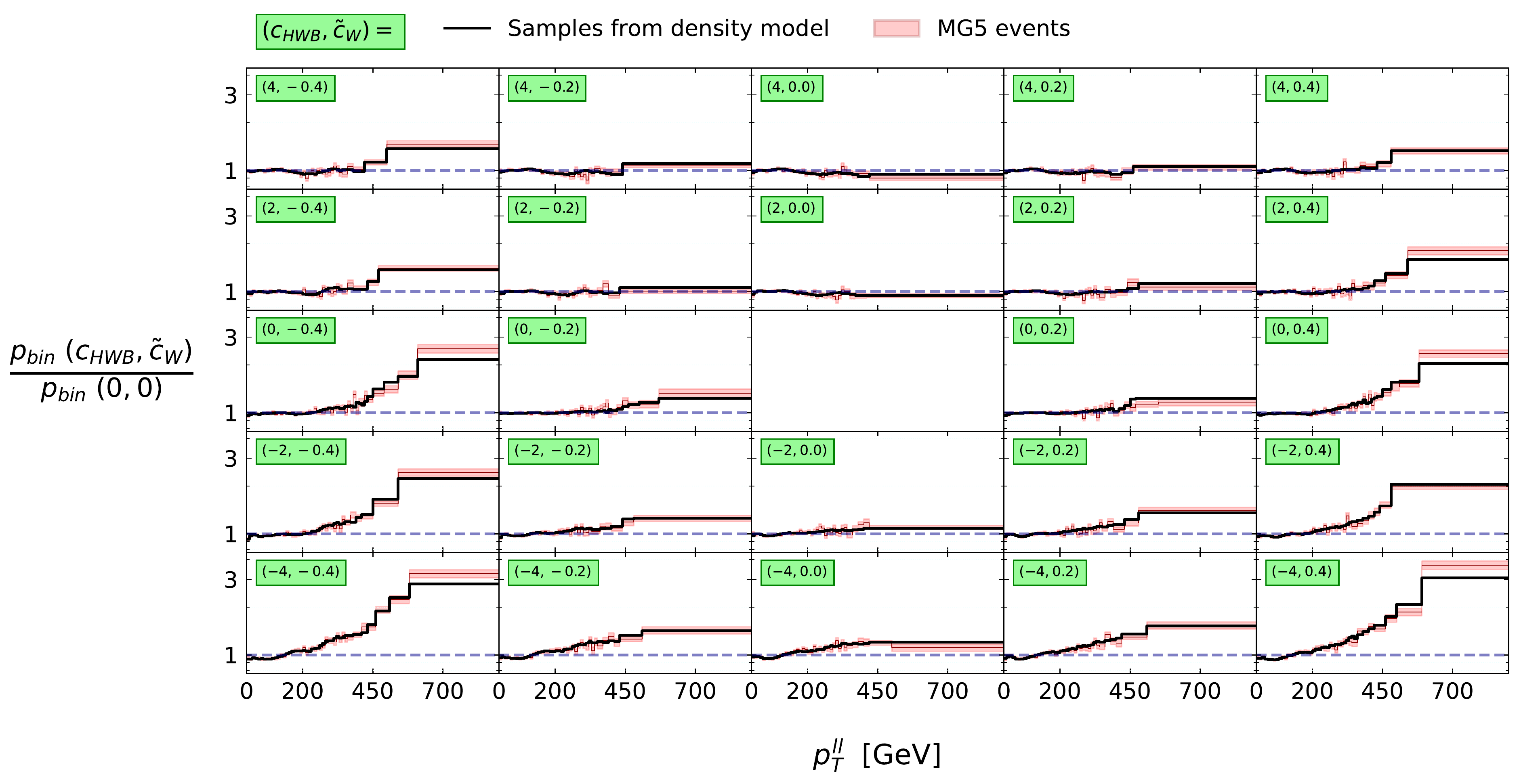}
    \caption{Evolution of the $p_\mathrm{T}^{ll}$ PDF as a function of $\left(c_\mathrm{HWB},{\tilde c}_W\right)$, presented as a ratio with respect to the SM hypothesis. The dependence is well captured by the density model. The double ratio comparing the two histograms is shown in Figure~\ref{fig:final-double-ratios-1}.}
    \label{fig:final-ratios-1}
    \vspace{1.0cm}
    \includegraphics[width=\textwidth]{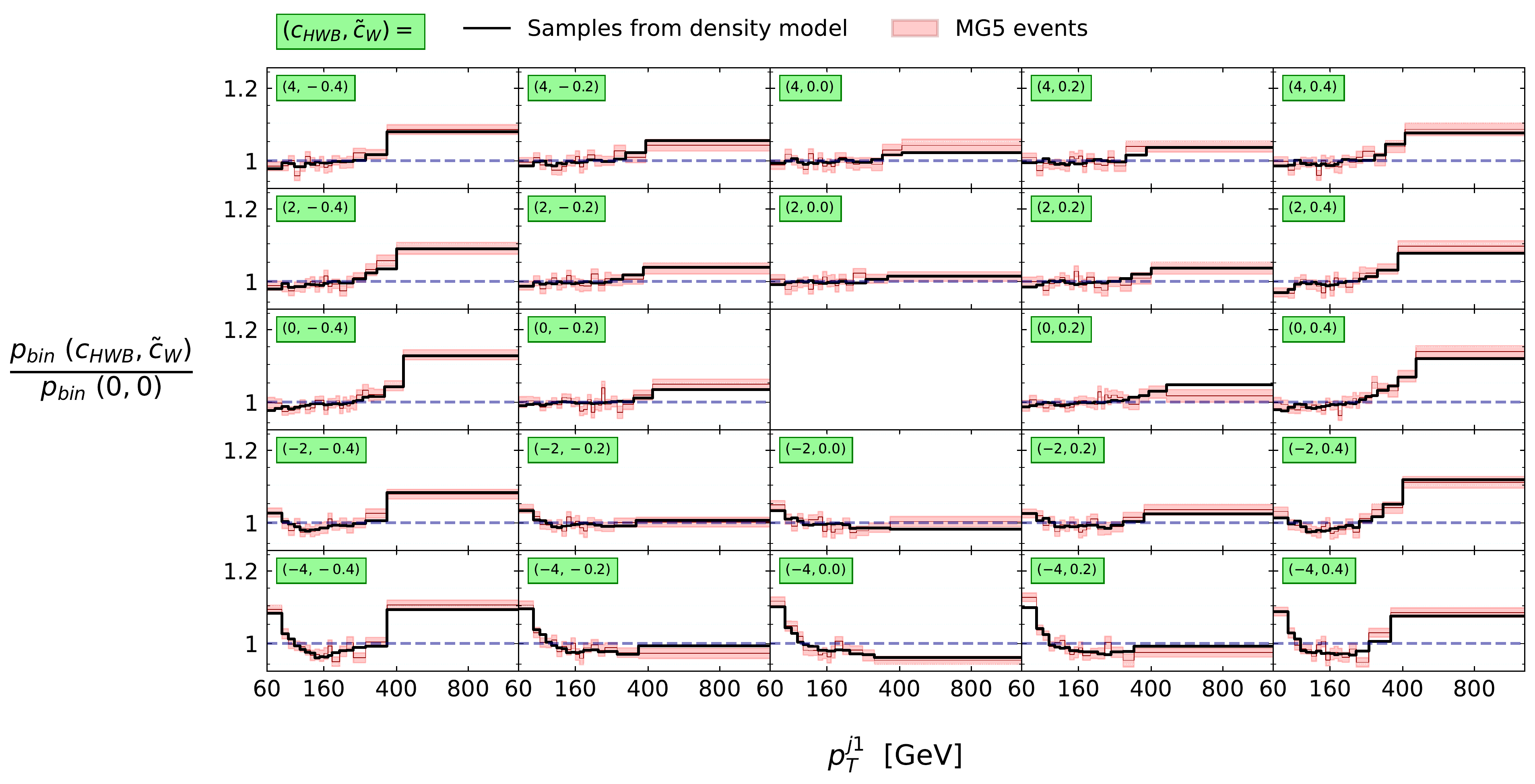}
    \caption{Evolution of the $p_\mathrm{T}^{j1}$ PDF as a function of $\left(c_\mathrm{HWB},{\tilde c}_W\right)$, presented as a ratio with respect to the SM hypothesis. The dependence is well captured by the density model. The double ratio comparing the two histograms is shown in Figure~\ref{fig:final-double-ratios-2}.}
    \label{fig:final-ratios-2}
\end{figure}

Figure~\ref{fig:final-ratios-1} shows how the $p_\mathrm{T}^{ll}$ PDF, expressed as a ratio with respect to the SM, varies as a function of the $\vec c$ hypothesis which is indicated by the green box in every panel. Events generated with \texttt{MG5} are shown in red, and those sampled from the density model are shown in black. We observe a significant enhancement of the high energy tail when ${\tilde c}_W$ is large in magnitude, approximately independent of its sign. We observe that negative values of $c_\mathrm{HWB}$ lead to a modest enhancement of the tail, whilst positive values suppress the tail by a comparable factor. The combination of these effects, plus any interference between them, manifests as a non-trivial structure throughout the plane of $\vec c$. We observe that the density model has captured this external parameter dependence well, since it is able to describe the deformations with an accuracy significantly better than the size of the deformations themselves. The double ratio, quantitatively comparing the two histograms, is presented in Figure~\ref{fig:final-double-ratios-1} of \ref{app:VBFZ-4obs-2param}.

Figure~\ref{fig:final-ratios-2} shows how the $p_\mathrm{T}^{j1}$ PDF varies as a function of $\vec c$. We observe an enhancement of the high-energy tail when ${\tilde c}_W$ is large in magnitude. We also observe a low-energy enhancement when $c_\mathrm{HWB}$ is highly negative, resulting in another non-trivial structure as we scan the plane of $\vec c$. Once again, we find that the density model has captured this external parameter dependence well. The double ratio, quantitatively comparing the two histograms, is presented in Figure~\ref{fig:final-double-ratios-2} of \ref{app:VBFZ-4obs-2param}.

\begin{figure}[p!]
    \includegraphics[width=\textwidth]{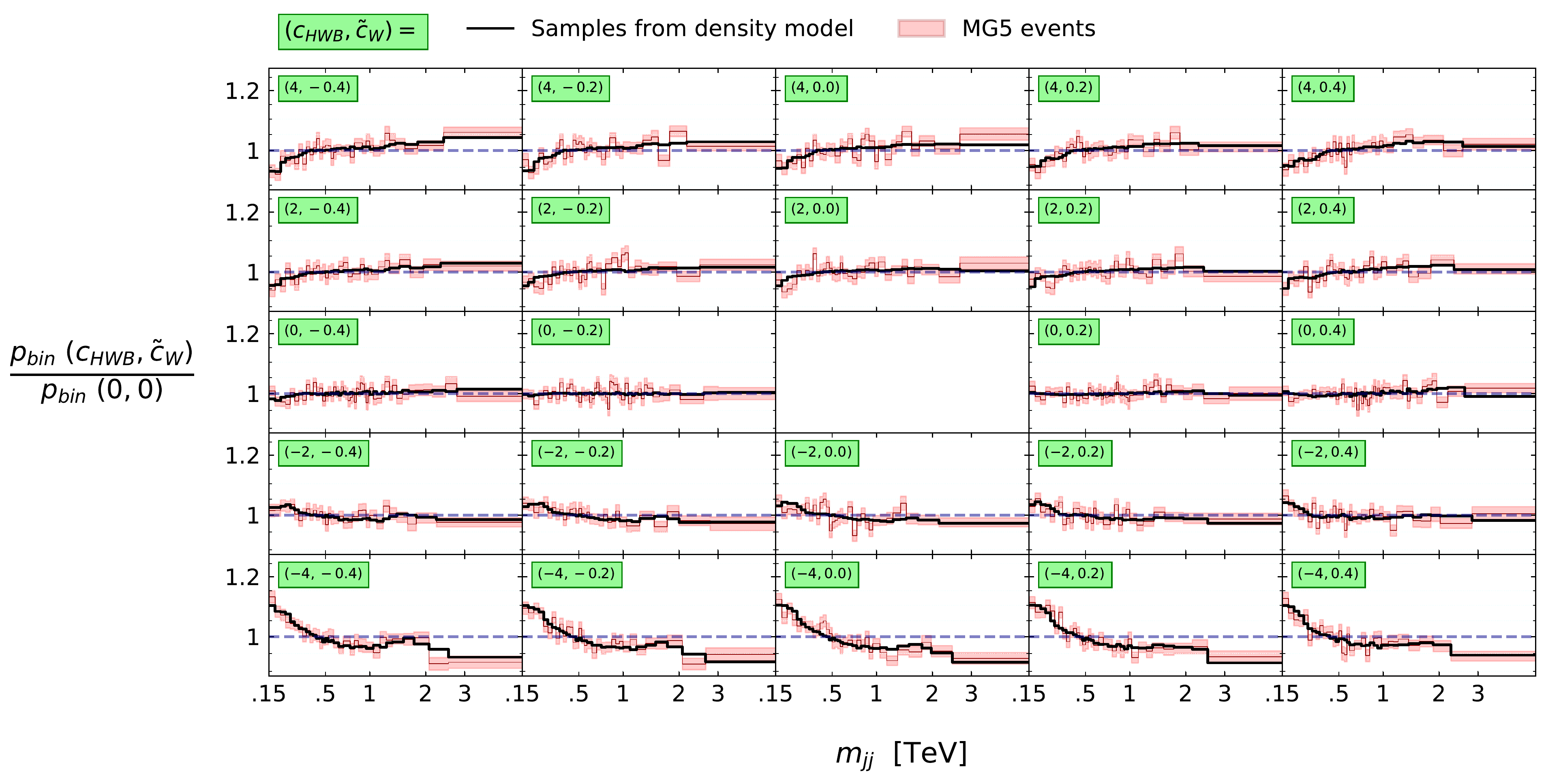}
    \caption{Evolution of the $m_{jj}$ PDF as a function of $\left(c_\mathrm{HWB},{\tilde c}_W\right)$, presented as a ratio with respect to the SM hypothesis. The dependence is well captured by the density model. The double ratio comparing the two histograms is shown in Figure~\ref{fig:final-double-ratios-3}.}
    \label{fig:final-ratios-3}
    \vspace{1.0cm}
    \includegraphics[width=\textwidth]{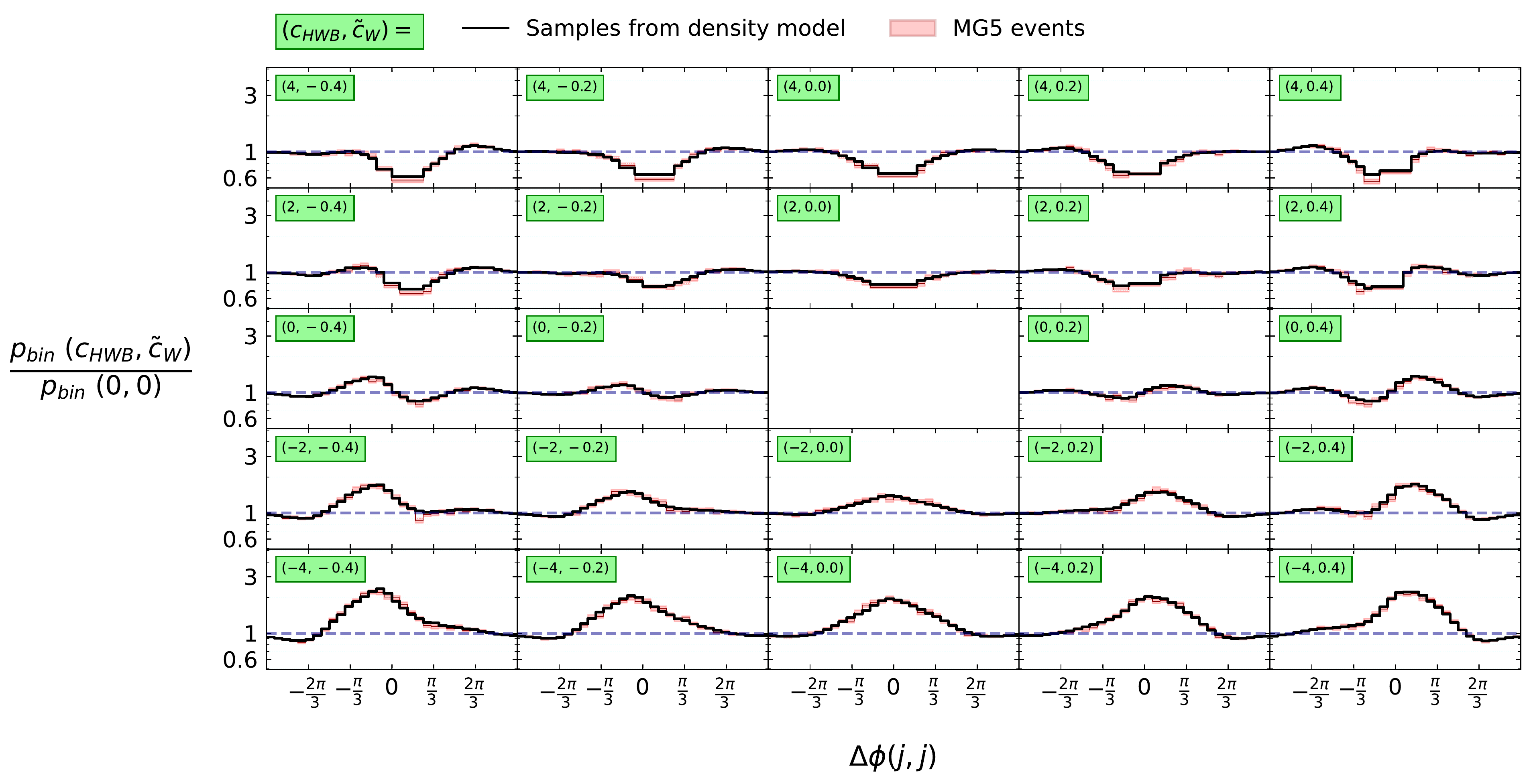}
    \caption{Evolution of the $\Delta\phi\left(j,j\right)$ PDF as a function of $\left(c_\mathrm{HWB},{\tilde c}_W\right)$, presented as a ratio with respect to the SM hypothesis. The dependence is well captured by the density model. The double ratio comparing the two histograms is shown in Figure~\ref{fig:final-double-ratios-4}.}
    \label{fig:final-ratios-4}
\end{figure}

Figure~\ref{fig:final-ratios-3} shows how the $m_{jj}$ PDF varies as a function of $\vec c$. We observe that highly negative values of $c_\mathrm{HWB}$ lead to significant structure at $m_{jj} \sim 0.15~\mathrm{TeV}$. As shown in Figure~\ref{fig:VBFZ-4obs-1D-dist}, this is also where the bulk of the data is expected to be measured. When measuring other observables, experimental analyses typically apply pre-selection criteria requiring $m_{jj}$ to exceed $\mathcal{O}\left(1~\mathrm{TeV}\right)$ in order to preferentially reject non-electroweak processes. By instead modelling an inclusive range of $m_{jj}$ simultaneously with all other observables and performing a high-dimensional unbinned analysis, such a restrictive requirement would not be required, provided that all backgrounds can also be sufficiently well modelled. The double ratio, quantitatively comparing the two histograms, is presented in Figure~\ref{fig:final-double-ratios-3} of \ref{app:VBFZ-4obs-2param}.

Figure~\ref{fig:final-ratios-4} shows how the $\Delta\phi\left(j,j\right)$ PDF varies as a function of $\vec c$. We observe that ${\tilde c}_W$ modulates the amplitude of an approximately sinusoidal oscillation introduced into the $\Delta\phi\left(j,j\right)$ spectrum. We observe that negative values of $c_\mathrm{HWB}$ modulate an enhancement at $\Delta\phi\left(j,j\right)\sim0$, whereas positive values of $c_\mathrm{W}$ cause a suppression. This observable is therefore sensitive to the sign of both parameters. Once again we note that the distribution shows a significantly non-trivial dependence as a function of $\vec c$, and that this dependence is captured well by the model. The double ratio, quantitatively comparing the two histograms, is presented in Figure~\ref{fig:final-double-ratios-4} of \ref{app:VBFZ-4obs-2param}.

\FloatBarrier
\section{Demonstration of statistical interpretation using a toy model}
\label{sec:toy-model-inference}

In the previous two sections we have demonstrated that we can construct density models which replicate the behaviour of simulated training data when sampled. Whilst this implies that good behaviour should also be obtained when performing inference tasks at the trained points in parameter space, this cannot be demonstrated because we are not able to evaluate the ground truth PDF for any given datapoint.

Nonetheless, we consider such a demonstration to be important. This is because the quality of inference is impacted not only by the ability to fit the training data 
but by (i) the degree of under- or over-training and (ii) the way in which the probability distribution is interpolated between training points, hereafter referred to as the inductive bias. Whilst the probability distribution may be learned with arbitrarily high accuracy \textit{at} the training points, depending on the complexity of the model configuration and number of training samples provided, it is likely that the interpolation between training points will not exactly match the true behaviour, which is unobserved. We aim to show that the approximate behaviour of the model can work sufficiently well for inference tasks, provided that training data are provided at dense enough points in parameter space.

To achieve this, we construct a toy model from which to sample ground truth training data. This is projected onto a latent space and used to train a density model using the method proposed in this paper. The toy contains four observables which vary according to two external parameters. Several pseudo-datasets are sampled from the true model assuming different parameter hypotheses. For each dataset, the density model is used to compute exclusion bounds on the latent space, and the results are compared with ground truth exclusion bounds computed using the true PDF on the data space. The level of agreement is then analyzed. Use of a toy model allows us to compute these ground truth bounds, which are typically intractable for real simulations.

We define a toy model with four observables $x ~=~ \{x_0,~x_1,~x_2,~x_3\}$ and two external parameters ${\vec c} ~=~ \{c_x,~c_y\}$. These observables are defined over the intervals $x_0 ~\in~ [100, ~800]$, $x_1 ~\in~ [100, ~800]$, $x_2 ~\in~ [-\pi, ~\pi]$ and $x_3 ~\in~ [-\infty,~\infty]$. \ref{app:toy-model-inference} defines the ground truth PDF and documents how samples are drawn. $50k$ datapoints are sampled at each of the $49$ parameter points in a two-dimensional grid spanning all permutations with $c_x \in \left[-1.5, -1, -0.5, 0, 0.5, 1, 1.5\right]$ and $c_y \in \left[-1.5, -1, -0.5, 0, 0.5, 1, 1.5\right]$.

\begin{figure}[htbp!]
    \centering
    \hspace{0.17cm} \includegraphics[width=0.792\textwidth]{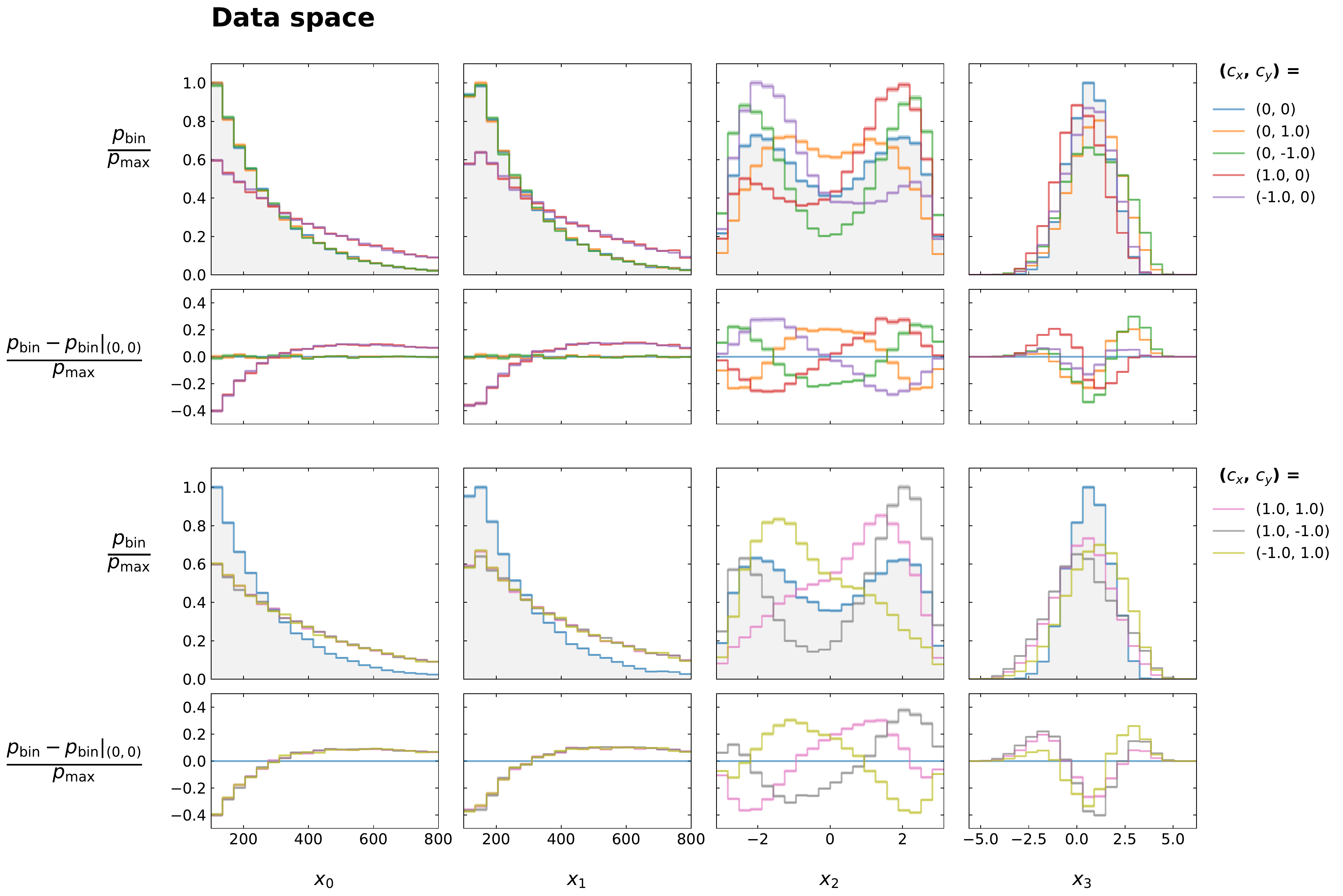} \\
    \vspace{0.3cm}
    \includegraphics[width=0.81\textwidth]{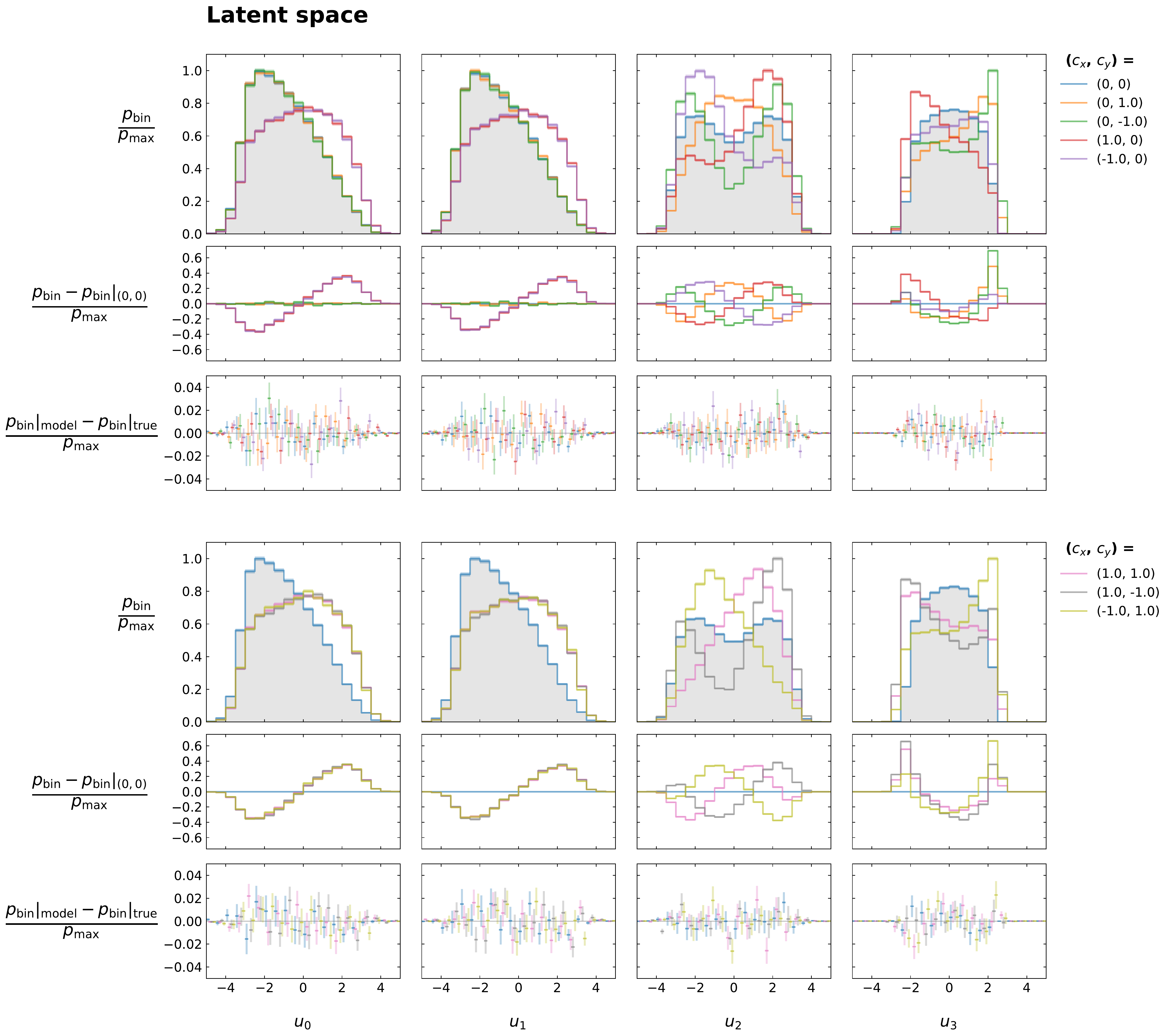}
    \caption{Kinematic distributions of toy model data before (top: ``Data space'') and after (bottom: ``Latent space'') projecting onto the latent space. Secondary panels highlight how these are modified by variations of the conditional parameters $\vec c = \left(c_x,~c_y\right)$. On the latent space, a third panel compares ground truth events with those sampled from the learned density model, demonstrating agreement within the statistical precision of the training data for all values of $\vec c$.}
    \label{fig:toy-1D-dists}
\end{figure}

Figure~\ref{fig:toy-1D-dists} (top) shows the 1D marginal distributions at the null hypothesis $\vec c=\left(0,0\right)$ as well as several alternative hypotheses in the $\vec c$-plane. Observables $x_0$ and $x_1$ are highly correlated falling distributions, where variations of $c_x$ away from $0$ enhance the amplitude in the tail. These observables are insensitive to $c_y$ as well as the sign of $c_x$. Observable $x_2$ is an angular observable for which $c_x$ and $c_y$ induce sinusoidal oscillations with a phase difference of $\frac{\pi}{2}$. This observable is sensitive to the sign and amplitude of both external parameters. Observable $x_3$ follows a smooth-peak distribution with no physical limits, and is correlated with all observables and external parameters.

Data are projected onto the latent space using values of $f=0.5$ for all observables. Neural networks are configured using the constants presented in Table~\ref{tab:toy-model-constants} and contain $18{\rm k}-85{\rm k}$ trainable parameters. 
Each network is trained on 60~\% of the available data until the log-likelihood evaluated over the other 40~\% no longer improves by an amount greater than $10^{-6}$ over a period of 8 consecutive epochs, after which the solution with the least-positive (or most-negative) validation loss is chosen. Training is found to terminate after $33-46$ epochs. Figure~\ref{fig:toy-1D-dists} (bottom) shows the latent space distributions. A third panel compares the the 1D marginal distributions obtained from the ground truth data and from drawing $50k$ samples from the resulting density model. The level of agreement is found to be comparable with the statistical precision of the data.
\begin{table}[htbp]
\caption{Constants used to construct and train a density model describing toy data with 4 observables and 2 external parameters.}
    \label{tab:toy-model-constants}
\begin{indented}
\item[]\begin{tabular}{@{}ccccc}
\br 
        $N_G=20$ & $A_1=50$ & $A_2=0$ & $B_1=50$ & $B_2=20$ \\

         $C=2$ & $D=3$ & $s_f=0.01$ & $s_\mu=0.01$ & $s_\sigma=0.01$ \\

        $f_\sigma=0.25$ & batch size = $500$ & $\lambda_{\rm lr}=0.001$  &  $\lambda_{\rm lr}^{\rm update~factor}=0.5$  &  $\lambda_{\rm lr}^{\rm patience}=2$ \\
\br
\end{tabular}
\end{indented}
\end{table}

\begin{figure}[htbp]
    \centering
    \subfigure[Model trained with nominal ${\vec c}_{\rm trained}$.]{\includegraphics[width=0.88\textwidth]{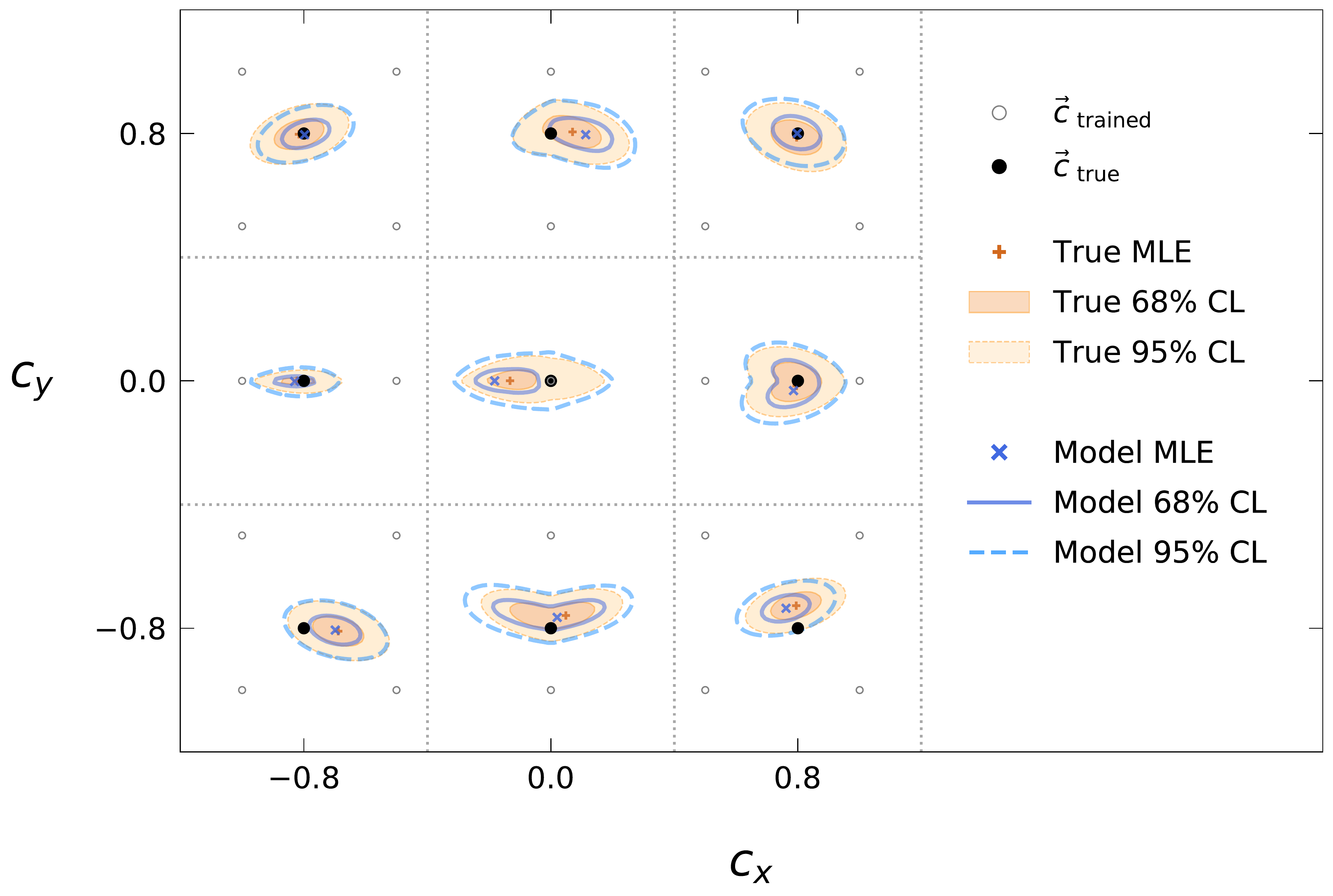}.}
    \subfigure[Model trained with additional ${\vec c}_{\rm trained}$ points.]{\includegraphics[width=0.88\textwidth]{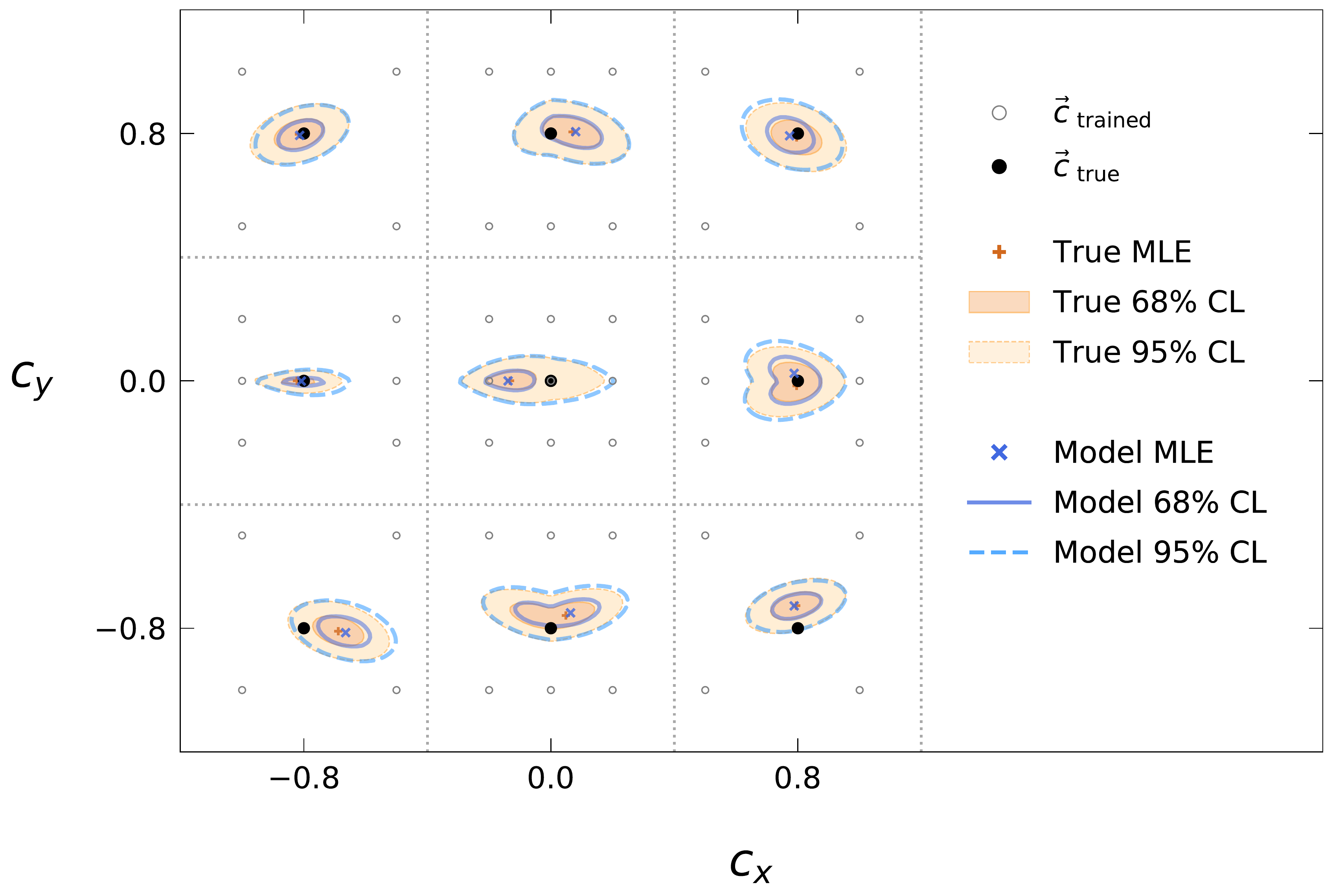}}
    \caption{68\% and 95\% confidence level contours in the $\vec c$-plane for nine separate datasets of size $N=400$ randomly sampled around the hypotheses ${\vec c}_{\rm true}$ shown in black. Contours are evaluated on the data space using the true probability model (orange) and on the latent space using the density model (blue). Crossed markers show the corresponding maximum likelihood estimators (MLEs). Good agreement is observed. Open circles show the points in parameter space ${\vec c}_{\rm trained}$ at which the model is trained.}
    \label{fig:toy-contours}
\end{figure}

We now test the accuracy of inference performed using the density model. We select nine different ``true'' hypotheses ${\vec c}_\mathrm{true}$ in a 2D grid with edges at $c_x \in \left[-0.8, 0, 0.8\right]$ and $c_y \in \left[-0.8, 0, 0.8\right]$. For each value of ${\vec c}_\mathrm{true}$, a pseudo-dataset with a size of $400$ events is created by sampling the true PDF. We assume that the expected number of observed events is identical for every value of ${\vec c}$. Figure~\ref{fig:toy-contours} (a) shows nine panels in which the different ${\vec c}_\mathrm{true}$ hypotheses are presented as black dots. Open circles show the points in parameter space ${\vec c}_{\rm trained}$ at which the model was trained, excluding those which lie outside of the axis range.

The true PDF is used to profile the likelihood of the dataset. Using this method we evaluate (i) the true maximum likelihood estimate (MLE) and (ii) the frequentist 68~\% and 95~\% confidence limits, assuming that the expected distribution of the profile likelihood ratio follows the asymptotic approximation described by Wilks' theorem\,\cite{wilks,wald}. In Figure~\ref{fig:toy-contours} (a), orange crosses present the MLE evaluated using the true PDF, whilst orange contours present the confidence limits. We note that, since the pseudo-datasets are stochastically sampled from the true PDF, we expect each MLE to fluctuate away from ${\vec c}_\mathrm{true}$ as observed. The datasets are then transformed onto the latent space, and the same analysis is performed using the density model to evaluate the likelihood. Blue crosses present the MLE evaluated using the density model, whilst blue contours present the confidence limits.

Figure~\ref{fig:toy-contours} (a) demonstrates generally good agreement between the exclusions bounds evaluated using the density model and ground truth PDF, although we observe a mild over-coverage when $c_x \sim 0$ or $c_y \sim 0$. We expect that this is because these axes represent turning points in the function $p\left(x|\vec c\right)$, the form of which is only approximated by the inductive bias of the density model. To test this, we train a second model which contains additional training data at $c_x=\pm0.2$ and $c_y=\pm0.2$. The resulting contours are shown in Figure~\ref{fig:toy-contours} (b). We observe that the additional training data have constrained the model at $|c_x|,|c_y| \sim 0$, resulting in an improved agreement with the ground truth. We conclude that the most reliable results will be achieved when the spacing of ${\vec c}_{\rm trained}$ points is smaller than the size of the expected exclusion bounds.

In both cases, Figure~\ref{fig:toy-contours} shows that accurate MLEs and exclusion contours have been estimated using density models on the latent space. Reliable results could therefore be obtained in this example without having access to the true PDF. 

\section{Conclusion}
\label{sec:conclusion}

We present a method for modelling probability distributions over a high-dimensional space of observables with dependence on external parameters, a dataset type which is common within the physical sciences. The method uses a novel transformation of input data and a targeted network architecture to improve the expressive power of Gaussian mixture models. It is designed to capture smooth deformations of the probability density induced by external parameter variations, and respects strict boundaries on the observables. The model may be used to perform inference on observed data, or sampled to act as a stochastic generator. 

We demonstrate the power of the method by applying it to two high-energy particle physics datasets: one which contains twelve highly correlated observables, and one which depends on two external parameters. We then use a toy model to demonstrate that fast and accurate inference may be performed from experimental data. We demonstrate that the problem-of-interest may also contain discrete observables, which are modelled with a relatively simple categorical model. 
Whilst the method enables interpretations to be performed using unbinned multi-dimensional data, it may also be used within the experimental design of binned measurements (which are intended to characterize observed data with minimal physical model assumptions). Such an analysis may proceed as follows. An experimenter may assign benchmark hypotheses to which a planned measurement should have reasonably optimized sensitivity. We expect that a near-optimal classifier\footnote{The Neyman-Pearson lemma states that the PDF ratio is the test-statistic with the highest fake rejection rate for a given true positive rate \cite{doi:10.1098/rsta.1933.0009}.} for a given parameter hypothesis may be created using the ratio of the PDFs evaluated at the null and alternative hypotheses. By isolating the regions of the high-dimensional space which provide the most discrimination power, they may ensure that these regions are targeted by dedicated bins.

The method presented is not domain-specific, and may be used to model any dataset of continuous observables which follow a smooth PDF, and to subsequently perform statistical inference from experimental data for the purposes of scientific discovery.

\section*{Data and code availability}
The code implementing the methodology developed in this paper and which can be used to reproduce these results is available at Ref.~\cite{CODE}. All corresponding simulated data and neural network model files are openly available at Ref.~\cite{DATA}

\begin{ack}
Darren Price is supported by a Turing Fellowship from the Alan Turing Institute, London, UK, by the Science and Technology Facilities Council (STFC) under grant ST/N000374/1, and by the University of Manchester. Stephen Menary is supported through a grant from the Alan Turing Institute and STFC grant ST/N000374/1.
\end{ack}

\appendix

\section{Double ratio plots comparing MG5 events with those sampled from the density model constructed used in Section~\ref{sec:VBFZ-4obs-2param}}
\label{app:VBFZ-4obs-2param}

\begin{figure}[p!]
    \includegraphics[width=\textwidth]{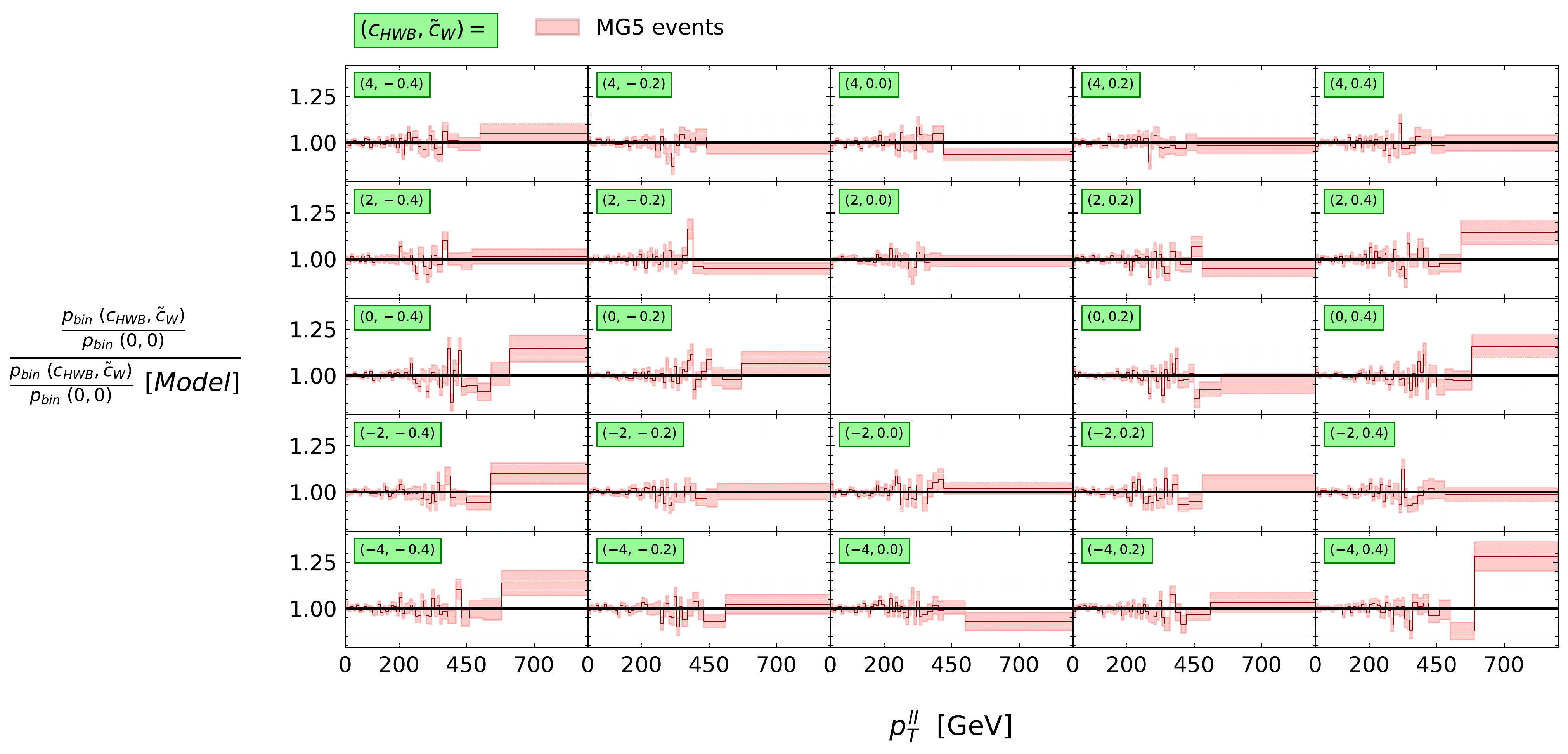}
    \caption{Double ratio comparing $p_{\mathrm bin}\left(c_\mathrm{HWB},{\tilde c}_W\right) ~/~ p_{\mathrm bin}\left(0,0\right)$ estimated using \texttt{MG5} events with those sampled from the density model, shown as a function of $p_\mathrm{T}^{ll}$. The corresponding single ratios are shown in Figure~\ref{fig:final-ratios-1}.}
    \label{fig:final-double-ratios-1}
    \vspace{1.0cm}
    \includegraphics[width=\textwidth]{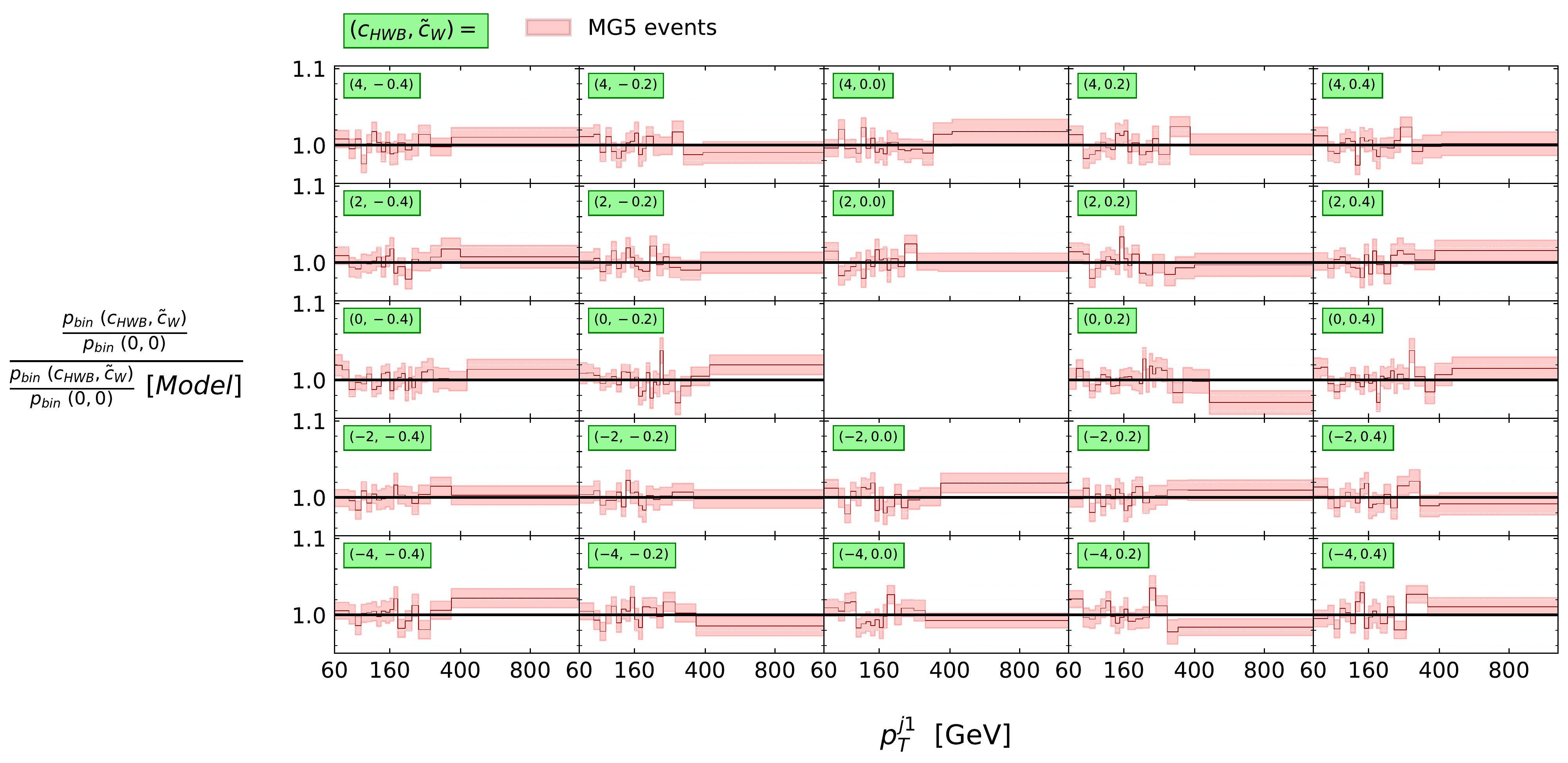}
    \caption{Double ratio comparing $p_{\mathrm bin}\left(c_\mathrm{HWB},{\tilde c}_W\right) ~/~ p_{\mathrm bin}\left(0,0\right)$ estimated using \texttt{MG5} events with those sampled from the density model, shown as a function of $p_\mathrm{T}^{j1}$. The corresponding single ratios are shown in Figure~\ref{fig:final-ratios-2}.}
    \label{fig:final-double-ratios-2}
\end{figure}

\begin{figure}[p!]
    \includegraphics[width=\textwidth]{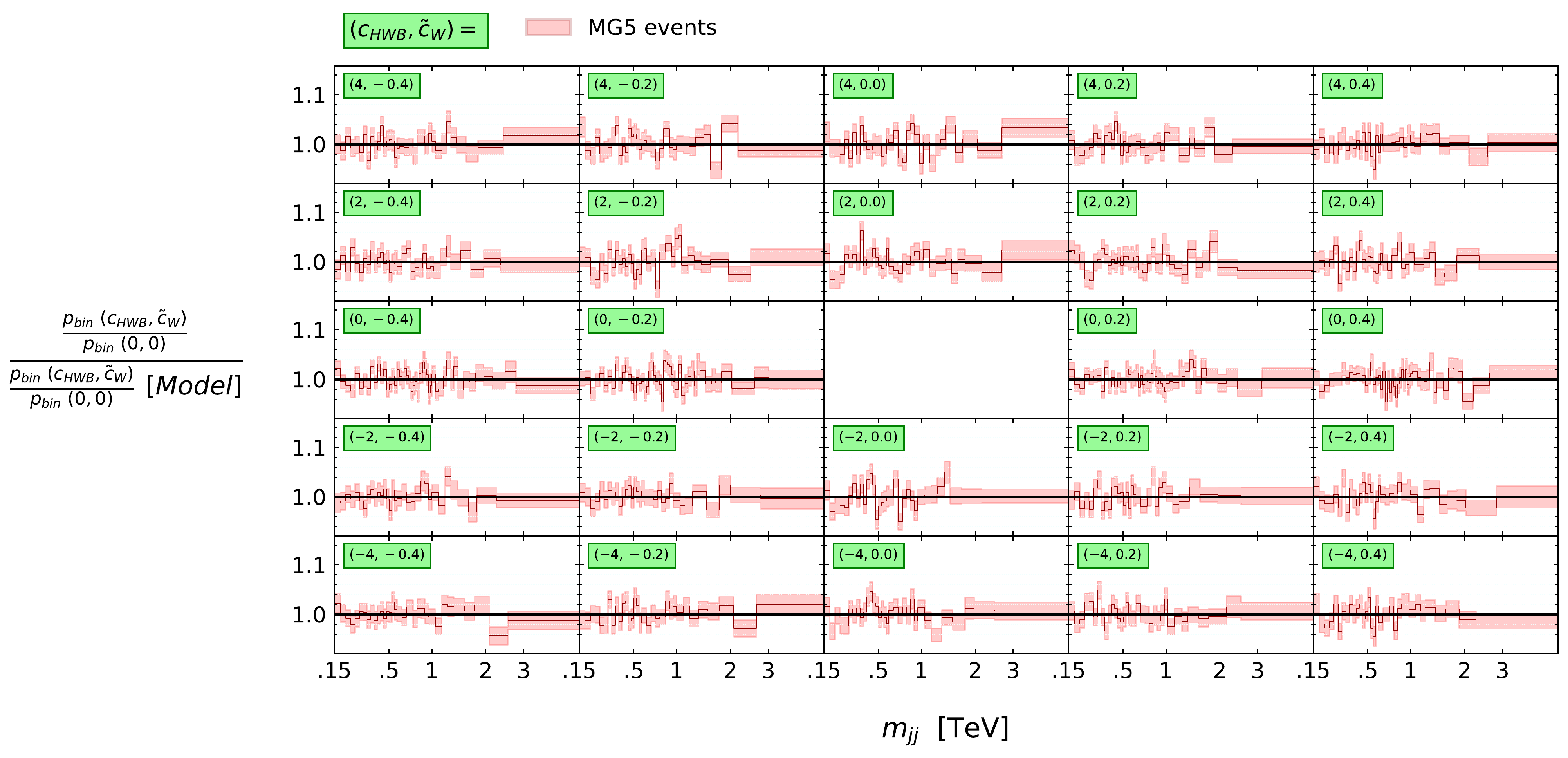}
    \caption{Double ratio comparing $p_{\mathrm bin}\left(c_\mathrm{HWB},{\tilde c}_W\right) ~/~ p_{\mathrm bin}\left(0,0\right)$ estimated using \texttt{MG5} events with those sampled from the density model, shown as a function of $m_{jj}$. The corresponding single ratios are shown in Figure~\ref{fig:final-ratios-3}.}
    \label{fig:final-double-ratios-3}
    \vspace{1.0cm}
    \includegraphics[width=\textwidth]{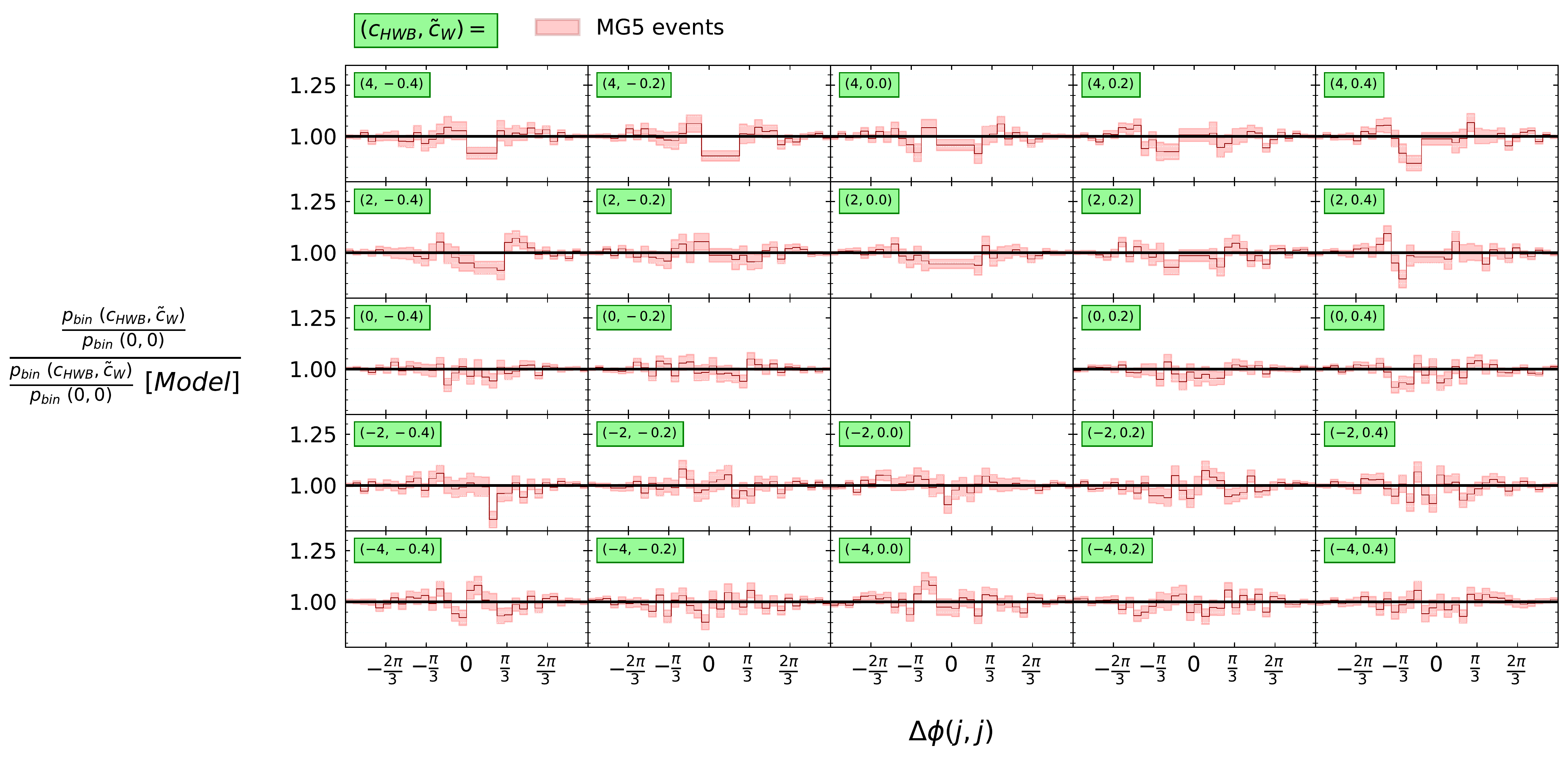}
    \caption{Double ratio comparing $p_{\mathrm bin}\left(c_\mathrm{HWB},{\tilde c}_W\right) ~/~ p_{\mathrm bin}\left(0,0\right)$ estimated using \texttt{MG5} events with those sampled from the density model, shown as a function of $\Delta\phi\left(j,j\right)$. The corresponding single ratios are shown in Figure~\ref{fig:final-ratios-4}.}
    \label{fig:final-double-ratios-4}
\end{figure}

This appendix presents further results concerning the experiments shown in Section~\ref{sec:VBFZ-4obs-2param}. In that section, Figures~\ref{fig:final-ratios-1}-\ref{fig:final-ratios-4} present the single-ratios comparing $p_{\mathrm bin}\left(c_\mathrm{HWB},{\tilde c}_W\right)$ with $p_{\mathrm bin}\left(0,0\right)$, visually demonstrating that the density model is able to capture deformations to the four observable spectra as $c_\mathrm{HWB}$ and ${\tilde c}_W$ are varied. Here, Figures~\ref{fig:final-double-ratios-1}-\ref{fig:final-double-ratios-4} show the double-ratios which quantitatively compare the single-ratios evaluated using $\texttt{MG5}$ events with the single-ratios evaluated using events sampled from the density model.

In Figures~\ref{fig:final-double-ratios-1}-\ref{fig:final-double-ratios-4} we observe that the double-ratio is consistent with unity at a level comparable with the estimated statistical uncertainty on the training data, which is presented as the red shaded area.

\FloatBarrier

\section{Ground truth probability density and sampling for the toy model used in Section~\ref{sec:toy-model-inference}}
\label{app:toy-model-inference}

For observables ${\vec x} ~=~ \{x_0,~x_1,~x_2,~x_3\}$ and external parameters ${\vec c} ~=~ \{c_x,~c_y\}$, the toy model described in Section~\ref{sec:toy-model-inference} is defined by a probability density
\begin{IEEEeqnarray}{lll}
    p_{\rm true}\left({\vec x}|{\vec c}\right) ~&=~ p_{\rm true}^{(0)}\left(x_0|c_x\right) ~\cdot~ p_{\rm true}^{(1)} \left(x_1|x_0\right) ~\cdot~ p_{\rm true}^{(2)}\left(x_2|{\vec c}\right) ~\cdot ~ p_{\rm true}^{(3)}\left(x_3|\vec c, x_1, x_2\right)
\end{IEEEeqnarray}

\noindent with the conditional probability densities
\begin{IEEEeqnarray*}{lll}
    p_{\rm true}^{(0)}\left(x_0|~c_x\right) ~&=~ \frac{1}{700} \cdot \frac{2(2-|c_x|)}{\left(1 - e^{-2\left(2-|c_x|\right)}\right)} \cdot e^{-2\left(2-|c_x|\right)\cdot x_0^{'}} \\
     p_{\rm true}^{(1)}\left(x_1|~x_0\right) ~&=~ \frac{1}{700} \cdot \frac{1}{\sqrt{\frac{\pi}{2}}\cdot\sigma_1\cdot\left({\rm erf}\frac{x_0^{'}}{\sqrt{2}\sigma_1}-{\rm erf}\frac{x_0^{'}-1}{\sqrt{2}\sigma_1}\right)} \cdot e^{-\frac{\left(x_1^{'} - x_0^{'}\right)^2}{2\cdot\sigma_1^2}} \IEEEyesnumber \\
     p_{\rm true}^{(2)}\left(x_2|~{\vec c}\right) ~&=~ \frac{\left(\alpha_2 + \beta_2 x_2^2 + \gamma_2 x_2^4\right) ~\cdot~ \left( 1 + \delta_2\left(c_x\right) \sin x_2 + \epsilon_2\left(c_y\right) \cos x_2 \right)}{f_2\left(\vec c, \pi\right) ~-~ f_2\left(\vec c, -\pi\right)} \\
     p_{\rm true}^{(3)}\left(x_3|~\vec c, x_1, x_2\right) ~&=~ q_3\left( x_3 ~+~ \frac{3}{5} \left(\sqrt{4 ~+~ |c_x|} ~+~ |c_y|\right) ~ \left(x_1^{'} ~+~ x_2^{'}\right) \right)
\end{IEEEeqnarray*}

\noindent defined over the intervals
\begin{IEEEeqnarray*}{lll}
 x_0 ~&\in~ [100, ~800] \\
 x_1 ~&\in~ [100, ~800] \\
 x_2 ~&\in~ [-\pi, ~\pi] \IEEEyesnumber \\
 x_3 ~&\in~ [-\infty,~\infty],
\end{IEEEeqnarray*}

\noindent where
\begin{IEEEeqnarray}{lll}
 x_0^{'} ~&=~ 2~\frac{x_0 ~-~ 100}{700}-1,\quad x_1^{'} ~=~ 2~\frac{x_1 ~-~ 100}{700}-1, \quad x_2^{'} ~=~ \frac{x_2 ~+~ \pi}{\pi}-1
\end{IEEEeqnarray}

\noindent with $\alpha_2 = 1$, $\beta_2=\frac{4}{\pi^2}$, $\gamma_2=-\frac{5}{\pi^4}$,  $\delta_2\left(c_x\right)=\frac{2}{5}c_x$, $\epsilon\left(c_y\right)=\frac{1}{2}c_y$, $\alpha_3=10$, $\beta_3=1$, $\gamma_3=1$ and 

\begin{IEEEeqnarray*}{rll}
    f_2\left(\vec c,~ x\right) ~&=~ \alpha_2 x ~+~ \frac{\beta_2}{3}x^3 ~+~ \frac{\gamma_2}{5}x^5 \\
    & \qquad +~ \big[\alpha_2\epsilon_2 ~+~ 2\beta_2\delta_2 x ~+~ \beta_2\epsilon_2\left(x^2-2\right) ~+~ 4\gamma_2\delta_2 x\left(x^2-6\right) \\
    & \qquad\qquad +~ \gamma_2\epsilon_2\left(x^4 ~-~ 12x^2 + 24\right)\big] \sin x \\
    & \qquad +~ \big[-\alpha_2\delta_2 ~+~ 2\beta_2\epsilon_2 x ~-~ \beta_2\delta_2\left(x^2-2\right) ~+~ 4\gamma_2\epsilon_2x\left(x^2-6\right) \\
    & \qquad\qquad -~ \gamma_2\delta_2\left(x^4 ~-~ 12x^2 ~+~ 24\right)\big] \cos x, \\[1ex]
    q_3 \left( x \right) ~&=~  \frac{1}{\left(1 + \exp[\alpha_3\left(x-\beta_3\right)-\gamma_3]\right)} \cdot \frac{1}{\left(1 + \exp[-\alpha_3\left(x-\beta_3\right)-\gamma_3]\right)} \cdot \frac{1}{2\left(\alpha_3\beta_3 + \gamma_3\right)f_3}, \\[1ex]
    f_3 ~&=~  \frac{1}{\alpha_3} \cdot \frac{\exp[2\left(\alpha_3\beta_3+\gamma_3\right)]}{ \exp[2\left(\alpha_3\beta_3+\gamma_3\right)] -1}, \IEEEyesnumber \\[1ex]
    g_3 ~&=~  f_3 \cdot \left( \alpha_3\beta_3 ~+~ \gamma_3\right), \\[1ex]
    h_3 \left(x\right) ~&=~  \exp \big[ \frac{g_3\left(2x - 1\right)}{f_3} \big].
\end{IEEEeqnarray*}

\noindent Samples are drawn according to:
\begin{IEEEeqnarray*}{lll}
    x_0^* ~&= 100 ~-~ 700 \cdot \frac{1}{2\left(2-c_x\right)} \cdot \log\left(1 - i_0^*\left(1 - e^{-2\left(2-|c_x|\right)}\right) \right) \\
    x_1^* ~&= 100 ~+~ 700 \big[x_0^{'} - \sqrt{2} \sigma_1 {\rm erf}^{-1}\left( \left(1-i_1^*\right) {\rm erf}\left(\frac{x_0^{'}}{\sqrt{2}\sigma_1}\right) ~+~ {\rm erf}\left(\frac{x_0^{'}-1}{\sqrt{2}\sigma_1}\right) i_1^* \right)\big] \\
    x_2^* ~&= I_2^{-1}\left(\vec c,~i_2^* \right) \IEEEyesnumber \\
    x_3^* ~&= I_3^{-1}\left( i_3^* \right) ~-~ \frac{3}{5} \left(\sqrt{4 ~+~ |c_x|} ~+~ |c_y|\right) \left(x_1^* ~+~ x_2^*\right)
\end{IEEEeqnarray*}

\noindent where $I_2^{-1}$ is evaluated numerically as the inverse function of
\begin{equation}
    I_2\left(\vec c,~x\right) ~=~ \frac{f_2\left(\vec c,~x\right) ~-~ f_2\left(\vec c,~-\pi\right)}{f_2\left(\vec c,~\pi\right) ~-~ f_2\left(\vec c,~-\pi\right)}
\end{equation}

\noindent and
\begin{equation}
    I_3^{-1}\left(i_3\right) ~=~ \frac{1}{\alpha_3} \log \frac{h_3\left(i_3\right)\exp\big[ \alpha_3\beta_3 ~+~ \gamma_3 \big] ~-~ 1}{\exp\big[ \alpha_3\beta_3 ~+~ \gamma_3 \big] ~-~ h_3\left(i_3\right)}.
\end{equation}

\section*{References}
\bibliographystyle{unsrt}
\bibliography{references}

\end{document}